\documentclass[conference]{IEEEtran}

\usepackage{graphicx}

\usepackage[T1]{fontenc}
\usepackage[utf8]{inputenc}

\usepackage{csquotes}
\usepackage{nicefrac}
\usepackage{booktabs}
\usepackage{tikz}
\usetikzlibrary{shapes.geometric, positioning, quantikz}
\usepackage{amssymb}
\usepackage{amsthm}

\usepackage[font=footnotesize]{subcaption}

\usepackage{yquant}
\useyquantlanguage{groups}
\usepackage{pgfplots}
\usepackage{flushend}

\usepackage[style=ieee, maxnames=6, minnames=1, date=year, doi=false,isbn=false,backend=biber]{biblatex}
\usepackage[binary-units=true, detect-all=true]{siunitx}

\usepackage{hyperref}
\newtheorem{example}{Example}

\AtEveryBibitem{
  \clearname{editor}%
  \clearfield{series}%
  \clearfield{isbn}%
  \clearfield{issn}%
  \clearfield{volume}%
  \clearfield{number}%
  \clearfield{pages}%
  \clearfield{doi}%
  \clearfield{urldate}
}

\DeclareFieldFormat
  [misc, book]
  {title}{\mkbibquote{#1}}

\begin{document}

\title{Reducing the Compilation Time of Quantum Circuits Using Pre-Compilation on the Gate Level}

\author{
	\IEEEauthorblockN{Nils Quetschlich\IEEEauthorrefmark{1}\hspace*{1.5cm}Lukas Burgholzer\IEEEauthorrefmark{1}\hspace*{1.5cm}Robert Wille\IEEEauthorrefmark{1}\IEEEauthorrefmark{2}}
	\IEEEauthorblockA{\IEEEauthorrefmark{1}Chair for Design Automation, Technical University of Munich, Germany}
	\IEEEauthorblockA{\IEEEauthorrefmark{2}Software Competence Center Hagenberg GmbH (SCCH), Austria}
	\IEEEauthorblockA{\href{mailto:nils.quetschlich@tum.de}{nils.quetschlich@tum.de}\hspace{1.5cm}\href{mailto:lukas.burgholzer@jku.at}{lukas.burgholzer@tum.de}\hspace{1.5cm} \href{mailto:robert.wille@tum.de}{robert.wille@tum.de}\\
	\url{https://www.cda.cit.tum.de/research/quantum}
	}
}

\maketitle

\begin{abstract}
In order to implement a quantum computing application, problem instances must be encoded into a quantum circuit and then compiled for a specific platform. 
The lengthy compilation process is a key bottleneck in this workflow, especially for problems that arise repeatedly with a similar yet distinct structure (each of which requires a new compilation run thus far).
In this paper, we aim to overcome this bottleneck by proposing a comprehensive \mbox{pre-compilation} technique that tries to minimize the time spent for compiling recurring problems while preserving the solution quality as much as possible.
The following concepts underpin the proposed approach:
Beginning with a problem class and a corresponding quantum algorithm, a \emph{predictive encoding} scheme is applied to encode a representative problem instance into a \mbox{general-purpose} quantum circuit for that problem class.
Once the real problem instance is known, the previously constructed circuit only needs to be adjusted---with (nearly) no compilation necessary.
Experimental evaluations on QAOA for the MaxCut problem as well as a case study involving a satellite mission planning problem show that the proposed approach significantly reduces the compilation time by several orders of magnitude compared to Qiskit's compilation schemes while maintaining comparable compiled circuit quality.
All implementations are available on GitHub (\url{https://github.com/cda-tum/mqt-problemsolver}) as part of the \emph{Munich Quantum Toolkit} (MQT).
\end{abstract}

\section{Introduction}\label{sec:intro}

Quantum computing is an emerging technology that is constantly improving both in software and hardware---sparking interest in academia and industry. 
As a result, many approaches have emerged that try to use quantum computing to solve problems from various application domains such as finance (e.g.,~\cite{stamatopoulosOptionPricingUsing2020}), machine learning (e.g.,~\cite{zoufalQuantumGenerativeAdversarial2019}), optimization (e.g.,~\cite{harwoodFormulatingSolvingRouting2021}), and chemistry (e.g.,~\cite{peruzzoVariationalEigenvalueSolver2014}).
Dedicated \emph{workflows} to describe the necessary steps from an initial problem to a solution using quantum computing are starting to emerge (e.g.,~\cite{quetschlich2023mqtproblemsolver, poggelRecommendingSolutionPaths2023, quetschlichPredictingGoodQuantum2023, quetschlich2023compileroptimization}).
These workflows usually include several steps: A suitable quantum algorithm has to be chosen for the respective problem and a quantum circuit needs to be constructed that encodes it.
Then, this circuit must be compiled and executed on a targeted quantum device before the solution can be decoded from the results of the execution.
Many \emph{Software Development Kits} (SDKs) have been developed for conducting these steps (or at least some of them).
Prominent examples are IBM's Qiskit~\cite{qiskit}, Quantinuum's TKET~\cite{tket}, Google's Cirq~\cite{cirq_developers_2023_8161252}, and BQSKit~\cite{osti_1785933}.
Additionally, there are quantum computing platform providers that offer access to various quantum devices such as AWS Braket~\cite{AmazonBraketPython2022}, and Microsoft's Azure Quantum~\cite{bradbenAzureQuantumDocumentation} which implicitly take care of the compilation when executing quantum circuits.

Compilation, i.e., making sure that a particular quantum circuit can actually be executed on a targeted device, is an essential part of these workflows that involves many computationally hard problems such as gate synthesis~\cite{gilesExactSynthesisMultiqubit2013, amyMeetinthemiddleAlgorithmFast2013, millerElementaryQuantumGate2011, zulehnerOnepassDesignReversible2018, niemannAdvancedExactSynthesis2020,
pehamDepthoptimalSynthesisClifford2023} or qubit routing/mapping~\cite{siraichiQubitAllocation2018, zulehnerEfficientMethodologyMapping2019, 
matsuoEfficientMethodQuantum2019,
willeMappingQuantumCircuits2019,
liTacklingQubitMapping2019,
pehamOptimalSubarchitecturesQuantum2023,
hillmichExploitingQuantumTeleportation2021,
zulehnerCompilingSUQuantum2019}.
It is crucial to perform this task as efficiently as possible, since the possible overhead introduced by compilation directly correlates with the quality of the compiled circuit. 
However, this takes time.
Time that in existing workflows is spent almost exclusively at \enquote{runtime}, i.e., at the moment the entire problem instance is known.
Furthermore, every single instance of a particular problem class is compiled from scratch---although its instances have a similar yet distinct structure and are frequently recurring, such as problems in scheduling~\cite{mohammadbagherpoor2021exploring} and finance~\cite{orus2019quantum}.

\begin{figure*}[t]
\centering
				\begin{subfigure}[c]{1.0\textwidth}
				\hspace{4.7cm}
				\begin{tikzpicture}
				\draw [decorate,decoration={brace,amplitude=10pt,mirror,raise=0pt}, rotate=90]
(0,8) -- (0,15.4) node [black,midway,above, yshift=3mm] {\footnotesize Compilation at runtime};
				\end{tikzpicture}
				\end{subfigure}
		\begin{subfigure}[c]{0.08\textwidth}
   				\resizebox{0.99\linewidth}{!}{
   				\begin{tikzpicture}
    \node[shape=circle,draw=black,minimum size=1.5cm] (B) at (0,3) {\huge 0};
    \node[shape=circle,draw=black,minimum size=1.5cm] (D) at (4,3) {\huge 1};
    \node[shape=circle,draw=black, minimum size=1.5cm] (A) at (0,0) {\huge 2};
    \node[shape=circle,draw=black,minimum size=1.5cm] (C) at (4,0) {\huge 3};

    \path [color=black, draw, line width =1mm] (A) edge node[] {}(B);
    \path [color=black, draw, line width =1mm]  (A) edge node[] {}(C);
    \path [color=black, draw, line width =1mm] (A) edge node[] {}(D);
    \path [color=black, draw, line width =1mm] (B) edge node[] {} (D);
   				\end{tikzpicture}
   				}
		\end{subfigure}
				$\xrightarrow{Enc.}$
				\begin{subfigure}[c]{0.1\textwidth}
   				\resizebox{0.99\linewidth}{!}{
   				\begin{tikzpicture}
   				
				  \begin{yquant}

						qubit {$q_0$} q;
						qubit {$q_1$} q[+1];
						qubit {$q_2$} q[+1];
						qubit {$q_3$} q[+1];
						
						[fill=purple!20]
				    	box {} (q[2],q[3]);
				    	[fill=green!20]
				    	box {} q[0];
				    	[fill=green!20]
				    	box {} (q[0],q[1]);
				    	[fill=green!20]
				    	box {} q[0];
				    	[fill=blue!50]
				    	box {} (q[1],q[3]);
				  \end{yquant}
				  \end{tikzpicture}
				}
		\end{subfigure}
				$\xrightarrow{Synth.}$
		\begin{subfigure}[c]{0.12\textwidth}
   				\resizebox{0.99\linewidth}{!}{
				   				\begin{tikzpicture}
				  \begin{yquant}

						qubit {$q_0$} q;
						qubit {$q_1$} q[+1];
						qubit {$q_2$} q[+1];
						qubit {$q_3$} q[+1];
						
				    	[fill=green!20]
				    	box {} q[3];
				    	[fill=green!20]
				    	box {} (q[2],q[3]);
				    	[fill=green!20]
				    	box {} q[3];
				    	[fill=green!20]
				    	box {} q[0];
				    	[fill=green!20]
				    	box {} (q[0],q[1]);
				    	[fill=green!20]
				    	box {} q[0];
				    	[fill=blue!50]
				    	box {} (q[1],q[3]);
				  \end{yquant}
				  \end{tikzpicture}
				}
		\end{subfigure}
				$\xrightarrow{Map.}$
		\begin{subfigure}[c]{0.16\textwidth}
   				\resizebox{0.99\linewidth}{!}{
				   				\begin{tikzpicture}
				  \begin{yquant}

						qubit {$q_0$} q;
						qubit {$q_1$} q[+1];
						qubit {$q_2$} q[+1];
						qubit {$q_3$} q[+1];
						
				    	[fill=green!20]
				    	box {} q[3];
				    	[fill=green!20]
				    	box {} (q[2],q[3]);
				    	[fill=green!20]
				    	box {} q[3];
				    	[fill=green!20]
				    	box {} q[0];
				    	[fill=green!20]
				    	box {} (q[0],q[1]);
				    	[fill=green!20]
				    	box {} q[0];
				    	swap {} (q[2],  q[3]);
				    	[fill=green!20]
				    	box {} (q[1],q[2]);
				  \end{yquant}
				  \end{tikzpicture}
				}
		\end{subfigure}
				$\xrightarrow{Meas.}$
		\begin{subfigure}[c]{0.12\textwidth}
   				\resizebox{0.99\linewidth}{!}{
\begin{tikzpicture}
    \begin{axis}
        [
        ybar,
        ymax=0.5,ymin=0,
        font=\LARGE,
        xticklabel style={rotate=90},
        ,ylabel=Relative Frequency
       ,xtick={0,1,2,3,4,5,6,7,8,9,10}
        ,xticklabels={$\ket{0000}$, , $\ket{0100}$,$\ket{0101}$,$\ket{0110}$,,$\ket{1111}$}
        ]
        \addplot coordinates
        {(0,0.04)(1,0.0)  (2,0.04) (3,0.4) (4,0.04)(5,0)  (6,0.04)};
    \end{axis}
    
	\node[align=center] at (1.5,-1.0) {\huge{$...$}};		
	\node[align=center] at (5.2,-1.0) {\huge{$...$}};		
	
\end{tikzpicture}
				}
		\end{subfigure}
				$\xrightarrow{Dec.}$
				\begin{subfigure}[c]{0.08\textwidth}
   				\resizebox{0.99\linewidth}{!}{
				\begin{tikzpicture}
    \node[shape=circle,draw=black,minimum size=1.5cm] (B) at (0,3) {\huge 0};
    \node[shape=circle,draw=black,minimum size=1.5cm] (D) at (4,3) {\huge 1};
    \node[shape=circle,draw=black, minimum size=1.5cm] (A) at (0,0) {\huge 2};
    \node[shape=circle,draw=black,minimum size=1.5cm] (C) at (4,0) {\huge 3};
    \node[shape=ellipse,draw=blue, dashed, line width=1mm, minimum height=6cm, minimum width=3cm] () at (0,1.5) {};

    \path [color=black, draw, line width =1mm] (A) edge node[] {}(B);
    \path [color=black, draw, line width =1mm]  (A) edge node[] {}(C);
    \path [color=black, draw, line width =1mm] (A) edge node[] {}(D);
    \path [color=black, draw, line width =1mm] (B) edge node[] {} (D);
				\end{tikzpicture}
				}
		\end{subfigure}
		
	\hspace{-7mm}		
	\subfloat[Problem instance.\label{fig:wf1_2} ]{\hspace{.16\linewidth}}
	\hspace{-2mm}
	\subfloat[Encoded circuit.\label{fig:wf1_3} ]{\hspace{.16\linewidth}}
	\hspace{3mm}
	\subfloat[Synth. circuit.\label{fig:wf1_4} ]{\hspace{.16\linewidth}}
	\hspace{6mm}
	\subfloat[Mapped circuit.\label{fig:wf1_5} ]{\hspace{.16\linewidth}}
	\hspace{12mm}
	\subfloat[Result.\label{fig:wf1_6} ]{\hspace{.10\linewidth}}
	\hspace{9mm}
	\subfloat[Solution.\label{fig:wf1_7} ]{\hspace{.10\linewidth}}
	
				\begin{subfigure}[t]{1.0\textwidth}
				\begin{tikzpicture}
					\draw[orange!60, line width=1mm] (1,0) -- (19,0) node[black, below, text width=3cm, align=center, midway]{\small{Runtime}};  
				\end{tikzpicture}
			\end{subfigure}
	\caption{Established quantum computing workflow from a (MaxCut) problem to its solution.}
	\label{fig:workflow}
\end{figure*}
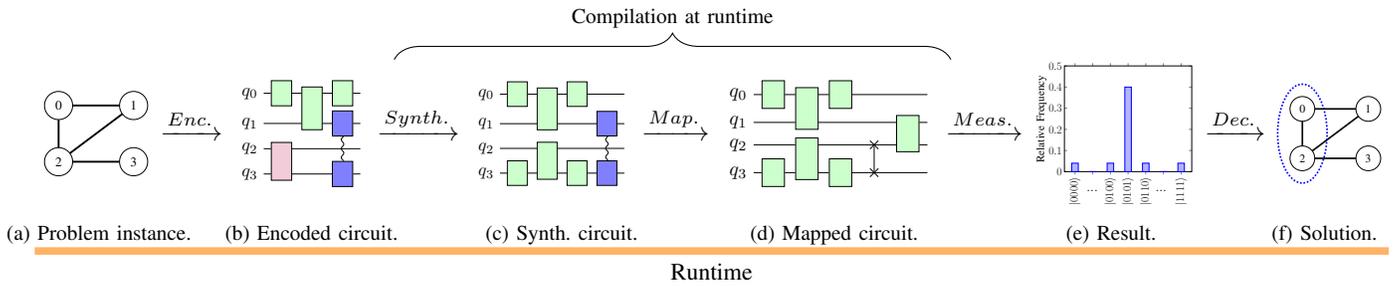  

In this work, we propose a \emph{pre-compilation} technique that aims to drastically reduce the time spent on compilation at runtime while maintaining comparable quality of the compiled circuits.
The proposed approach is based on the following ideas:
Starting with a problem class (e.g., the MaxCut problem) and a quantum algorithm to solve the problem (e.g., the \emph{Quantum Approximate Optimization Algorithm}, QAOA~\cite{farhi2014quantum}), a \emph{predictive encoding} scheme is applied that encodes the classes' structure (e.g., a QAOA circuit for fully connected graphs) into a \mbox{general-purpose} quantum circuit for the whole class of problems.
This circuit is then \mbox{(pre-) compiled} to yield a generic circuit for solving problems in the respective class that can be executed on a certain device.
At runtime, when the actual problem instance is known, the resulting circuit only needs to be adjusted for the particular instance (e.g., by removing gates corresponding to edges not present in the graph instance)---without requiring any compilation.
As the \mbox{general-purpose} circuits have to subsume a wide variety of instances, their compilation is likely to induce a higher overhead compared to just compiling one particular problem instance.
To counteract this potential loss in quality, some lightweight optimizations are performed at runtime in addition to adjusting the circuit.

\vspace{5cm}

Experimental evaluations on QAOA for the MaxCut problem as well as a \mbox{real-world} satellite mission planning \mbox{use-case} demonstrate the benefits of the proposed approach.
For instances ranging from $5$ to $100$ qubits, the compilation time at runtime is reduced by several orders of magnitudes with a similar compiled circuit quality compared to the established workflow with all implementations available on GitHub (\url{https://github.com/cda-tum/mqt-problemsolver}) as part of the \emph{Munich Quantum Toolkit} (MQT).

The remainder of this work is structured as follows:
The currently established quantum computing workflow from a problem to a solution is reviewed in \autoref{sec:flow}.
Subsequently, \autoref{sec:idea} motivates the proposed \mbox{pre-compilation} scheme and reviews related work.
Based on that, \autoref{sec:comp_scheme} describes the proposed compilation scheme in detail before it is evaluated in \autoref{sec:eval}.
Afterwards, the proposed scheme is applied to a real-world industry \mbox{use-case} from satellite mission planning in \autoref{sec:satellite} and discussed in \autoref{sec:discussion}.
Finally, \autoref{sec:conclusions} concludes this work.

\section{The Quantum Computing Workflow}\label{sec:flow}
Quantum computing can be used to solve classical problems from various domains as discussed in \autoref{sec:intro}.
Although the problems to be solved obviously vary, the general workflow from the problem to a solution is commonly structured as shown in \autoref{fig:workflow}.

Given a classical problem instance of a certain problem class, the first step to solve the problem with quantum computing is to \emph{encode} the problem into a quantum circuit that shall afterwards be executed on a quantum computer.
This encoding step is highly non-trivial and constitutes a whole research area on its own (e.g.,~\cite{dominguez2023encodingindependent,weigold2021encoding,schuld2021machine}).
It requires the problem description to be transformed in a way that is suitable as an input for a certain quantum algorithm---with many degrees of freedom in the choice of an encoding and a particular algorithm.

\begin{example}
Consider the graph shown in \autoref{fig:wf1_2} and assume that we want to solve the well-known MaxCut problem on this graph, i.e., the goal is to find a partition of the graph's nodes so that the number of edges between these partitions is maximal.
The \emph{Quantum Approximate Optimization Algorithm} (QAOA,~\cite{farhi2014quantum}) has been proposed as a candidate to tackle this problem on a quantum computer.
For this, each node is encoded as a qubit and each edge between two nodes is encoded as a certain interaction between the respective qubits---as sketched in \autoref{fig:wf1_3}.
\end{example}

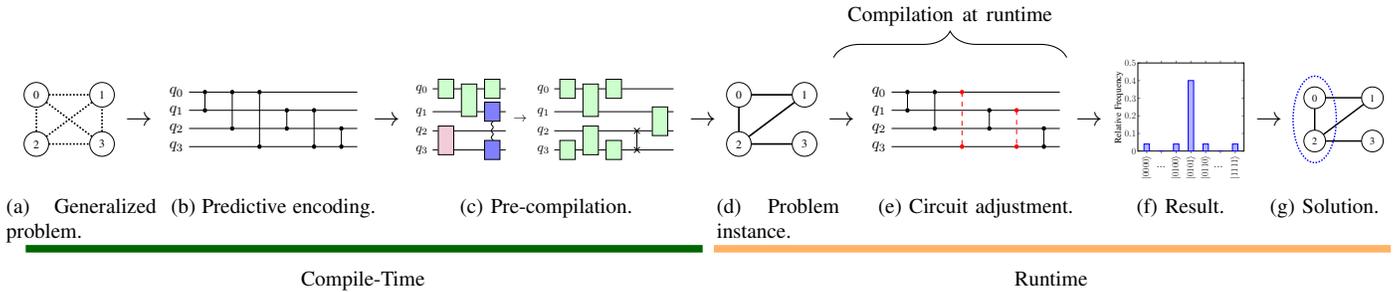
\begin{figure*}[t]
				\begin{subfigure}[c]{1.0\textwidth}
				\hspace{9.6cm}
				\begin{tikzpicture}
				\draw [decorate,decoration={brace,amplitude=10pt,mirror,raise=0pt}, rotate=90, yshift=8cm]
(8,8) -- (8,11.1) node [black,midway,above, yshift=3mm, text width=5cm, align=center] {\footnotesize Compilation at runtime};
				\end{tikzpicture}
				\end{subfigure}
				\begin{subfigure}[c]{0.07\textwidth}
   \resizebox{0.99\linewidth}{!}{
				\begin{tikzpicture}
    \node[shape=circle,draw=black,minimum size=1.5cm, ] (B) at (0,3) {\huge 0};
    \node[shape=circle,draw=black,minimum size=1.5cm] (D) at (4,3) {\huge 1};
    \node[shape=circle,draw=black, minimum size=1.5cm] (A) at (0,0) {\huge 2};
    \node[shape=circle,draw=black,minimum size=1.5cm] (C) at (4,0) {\huge 3};

    \path [color=black, dashed, line width =1mm] (A) edge node[] {}(B);
    \path [color=black, dashed, line width =1mm] (A) edge node[] {}(C);
    \path [color=black, dashed, line width =1mm] (A) edge node[] {}(D);
    \path [color=black, dashed, line width =1mm] (B) edge node[] {} (C);
    \path [color=black, dashed, line width =1mm] (B) edge node[] {} (D);
    \path [color=black, dashed, line width =1mm] (C) edge node[] {} (D); 
				\end{tikzpicture}
				}
				\end{subfigure}
				$\rightarrow$
				\begin{subfigure}[c]{0.15\textwidth}
   				\resizebox{0.99\linewidth}{!}{
   								\begin{tikzpicture}
				  \begin{yquant}

						qubit {$q_0$} q;
						qubit {$q_1$} q[+1];
						qubit {$q_2$} q[+1];
						qubit {$q_3$} q[+1];
					
				    	zz (q[0,1]);
				    	zz (q[0,2]);
				    	zz (q[0,3]);
				    	zz (q[1,2]);
				    	zz (q[1,3]);
				    	zz (q[2,3]);		
 				    	
				  \end{yquant}
				\end{tikzpicture}
   				}
				\end{subfigure}
				$\rightarrow$
				\begin{subfigure}[c]{0.2\textwidth}
   				\resizebox{0.99\linewidth}{!}{
   				\begin{tikzpicture}
				     \begin{yquantgroup}
      					\registers{
						qubit {$q_0$} q;
						qubit {$q_1$} q[+1];
						qubit {$q_2$} q[+1];
						qubit {$q_3$} q[+1];
      							}
				    	\circuit{
						[fill=purple!20]
				    	box {} (q[2],q[3]);
				    	[fill=green!20]
				    	box {} q[0];
				    	[fill=green!20]
				    	box {} (q[0],q[1]);
				    	[fill=green!20]
				    	box {} q[0];
				    	[fill=blue!50]
				    	box {} (q[1],q[3]);
				    	}
   				\equals[$\rightarrow$]

				\circuit{

				    	[fill=green!20]
				    	box {} q[3];
				    	[fill=green!20]
				    	box {} (q[2],q[3]);
				    	[fill=green!20]
				    	box {} q[3];
				    	[fill=green!20]
				    	box {} q[0];
				    	[fill=green!20]
				    	box {} (q[0],q[1]);
				    	[fill=green!20]
				    	box {} q[0];
				    	swap {} (q[2],  q[3]);
				    	[fill=green!20]
				    	box {} (q[1],q[2]);
				    	}

				  \end{yquantgroup}
				  \end{tikzpicture}
				}
		\end{subfigure}
				$\rightarrow$
				\begin{subfigure}[c]{0.07\textwidth}
   				\resizebox{0.99\linewidth}{!}{
   				\begin{tikzpicture}
    \node[shape=circle,draw=black,minimum size=1.5cm] (B) at (0,3) {\huge 0};
    \node[shape=circle,draw=black,minimum size=1.5cm] (D) at (4,3) {\huge 1};
    \node[shape=circle,draw=black, minimum size=1.5cm] (A) at (0,0) {\huge 2};
    \node[shape=circle,draw=black,minimum size=1.5cm] (C) at (4,0) {\huge 3};

    \path [color=black, draw, line width =1mm] (A) edge node[] {}(B);
    \path [color=black, draw, line width =1mm]  (A) edge node[] {}(C);
    \path [color=black, draw, line width =1mm] (A) edge node[] {}(D);
    \path [color=black, draw, line width =1mm] (B) edge node[] {} (D);
   				\end{tikzpicture}
   				}
		\end{subfigure}
				$\rightarrow$
		\begin{subfigure}[c]{0.15\textwidth}
   				\resizebox{0.99\linewidth}{!}{
				\begin{tikzpicture}
				  \begin{yquant}

						qubit {$q_0$} q;
						qubit {$q_1$} q[+1];
						qubit {$q_2$} q[+1];
						qubit {$q_3$} q[+1];
					
						[style=black]
				    	zz (q[0,1]);
						[style=black]
				    	zz (q[0,2]);
						[style=red, control style = dashed]
				    	zz (q[0,3]);
						[style=black]
				    	zz (q[1,2]);
						[style=red, control style = dashed]
				    	zz (q[1,3]);
						[style=black]
				    	zz (q[2,3]);		
 				    	
				  \end{yquant}
				\end{tikzpicture}
   				}
		\end{subfigure}
				$\rightarrow$
		\begin{subfigure}[c]{0.10\textwidth}
   				\resizebox{0.99\linewidth}{!}{
\begin{tikzpicture}
    \begin{axis}
        [
        ybar,
        ymax=0.5,ymin=0,
        font=\LARGE,
        xticklabel style={rotate=90},
        ,ylabel=Relative Frequency
       ,xtick={0,1,2,3,4,5,6,7,8,9,10}
        ,xticklabels={$\ket{0000}$, , $\ket{0100}$,$\ket{0101}$,$\ket{0110}$,,$\ket{1111}$}
        ]
        \addplot coordinates
        {(0,0.04)(1,0.0)  (2,0.04) (3,0.4) (4,0.04)(5,0)  (6,0.04)};
    \end{axis}
    
	\node[align=center] at (1.5,-1.0) {\huge{$...$}};		
	\node[align=center] at (5.2,-1.0) {\huge{$...$}};		
	
\end{tikzpicture}
				}
		\end{subfigure}
				$\rightarrow$
				\begin{subfigure}[c]{0.07\textwidth}
   				\resizebox{0.99\linewidth}{!}{
				\begin{tikzpicture}
    \node[shape=circle,draw=black,minimum size=1.5cm] (B) at (0,3) {\huge 0};
    \node[shape=circle,draw=black,minimum size=1.5cm] (D) at (4,3) {\huge 1};
    \node[shape=circle,draw=black, minimum size=1.5cm] (A) at (0,0) {\huge 2};
    \node[shape=circle,draw=black,minimum size=1.5cm] (C) at (4,0) {\huge 3};
    \node[shape=ellipse,draw=blue, dashed, line width=1mm, minimum height=6cm, minimum width=3cm] () at (0,1.5) {};

    \path [color=black, draw, line width =1mm] (A) edge node[] {}(B);
    \path [color=black, draw, line width =1mm]  (A) edge node[] {}(C);
    \path [color=black, draw, line width =1mm] (A) edge node[] {}(D);
    \path [color=black, draw, line width =1mm] (B) edge node[] {} (D);
				\end{tikzpicture}
				}
		\end{subfigure}
		
	\hspace{-3mm}
	\subfloat[Generalized problem.\label{fig:wf2_1}]{\hspace{.11\linewidth}}
	\hspace{1mm}
	\subfloat[Predictive encoding.\label{fig:wf2_2}]{\hspace{.15\linewidth}}
	\subfloat[Pre-compilation.\label{fig:wf2_3}]{\hspace{.25\linewidth}}
	\subfloat[Problem instance.\label{fig:wf2_4}]{\hspace{.09\linewidth}}
	\subfloat[Circuit adjustment.\label{fig:wf2_5}]{\hspace{.20\linewidth}}
	\subfloat[Result.\label{fig:wf2_6}]{\hspace{.10\linewidth}}
	\hspace{1mm}
	\subfloat[Solution.\label{fig:wf2_7}]{\hspace{.09\linewidth}}
	
				\begin{subfigure}{0.5\textwidth}
				\begin{tikzpicture}
					\draw[black!60!green,  line width=1mm] (0,0) -- (9,0);
				\end{tikzpicture}
				\caption*{Compile-Time}
			\end{subfigure}
				\begin{subfigure}{0.5\textwidth}
				\begin{tikzpicture}
					\draw[orange!60, line width=1mm] (0,0) -- (9,0);
				\end{tikzpicture}
				\caption*{Runtime}
			\end{subfigure}	
	
	\caption{Proposed workflow including pre-compilation.}
	\label{fig:workflow2}
	\vspace{-4mm}
\end{figure*} 

Once encoded as a quantum circuit, the circuit must be \emph{compiled} to be executable on a particular quantum device.
This is similar to classical programming, where a \mbox{high-level} program (e.g., written in C++) needs to be compiled into \mbox{low-level} machine instructions that are supported by the CPU on which the program should run.
Quantum computers generally offer a small but universal set of gates that is natively supported by the device.
In the following, this is referred to as \emph{native gate-set}.
Every quantum computation that shall be executed on a certain device must first be broken down into sequences of native gates---a process frequently referred to as \emph{synthesis} with many respective methods proposed, e.g., in~\cite{gilesExactSynthesisMultiqubit2013, amyMeetinthemiddleAlgorithmFast2013, millerElementaryQuantumGate2011, zulehnerOnepassDesignReversible2018, niemannAdvancedExactSynthesis2020, pehamDepthoptimalSynthesisClifford2023}.

\begin{example}
Consider again the circuit from \autoref{fig:wf1_3} and assume that one of its five gates (the one marked in red) is not a native gate on the targeted quantum device.
Then, \autoref{fig:wf1_4} sketches what a synthesized circuit might look like, where that gate has been decomposed into three (native) gates (now marked in green).
\end{example}

Now that the circuit consists only of native gates, the circuit's (logical) qubits need to be \emph{mapped} to the device's (physical) qubits---a process frequently referred to as \emph{qubit placement}, \emph{qubit allocation}, or \emph{qubit layout}.
Many quantum computers, such as those based on superconducting qubits or neutral atoms, have a limited connectivity between their qubits, i.e., \mbox{multi-qubit} gates may only be applied to qubits that are connected on the device.
As a result, the qubit mapping generally has to be adapted dynamically throughout the circuit.
This is typically achieved by inserting \emph{SWAP} operations in a process referred to as \emph{qubit routing}.
The whole process of determining an initial qubit placement and then routing the circuit is commonly referred to as \emph{mapping} with methods proposed, e.g., in~\cite{siraichiQubitAllocation2018, zulehnerEfficientMethodologyMapping2019, 
matsuoEfficientMethodQuantum2019,
willeMappingQuantumCircuits2019,
liTacklingQubitMapping2019,
pehamOptimalSubarchitecturesQuantum2023,
hillmichExploitingQuantumTeleportation2021,
zulehnerCompilingSUQuantum2019}.

\vspace{5cm}

\begin{example}
	Consider the synthesized circuit shown in \autoref{fig:wf1_4} and assume that each of the qubits is connected to its neighbors on the targeted device.
	Then, all but one gate (the one marked in blue) already adhere to the connectivity constraints.
	To resolve the remaining conflict, a single \emph{SWAP} gate is inserted in front of the last gate as shown in \autoref{fig:wf1_5}.
\end{example}

The circuit is now executable on the targeted device---marking the end of the compilation step considered in this work\footnote{Before a circuit can really be executed on a particular device, several more low-level steps such as pulse-level compilation and scheduling need to be conducted (using SDKs such as proposed in, e.g.,~\cite{Alexander_2020, Li2022pulselevelnoisy, Silverio2022pulseropensource}). These steps will not be considered in further detail in this work and are assumed to occur once the circuit is sent to the device provider for execution.}.
It is essential to keep the overhead introduced by compiling the original circuit as low as possible, since each added gate decreases the fidelity of the overall result.
The complexity of the underlying problems (e.g., mapping being NP-complete~\cite{Botea2018OnTC}) makes compilation a time- and resource-intensive task that is, nevertheless, crucial for reliably executing quantum circuits.

For the actual execution, the circuit is usually sent to a quantum device provider over the cloud and the result is sent back in the form of a histogram of measurement outcomes.
These measurement outcomes are then \mbox{post-processed} (\emph{decoded}) to reveal the solution to the problem.

\begin{example}
	An execution of the quantum circuit sketched in \autoref{fig:wf1_5} might yield a distribution of measurement outcomes as shown in \autoref{fig:wf1_6}.
	There, the measurement result ($\ket{0101}$) occurred with a significantly higher frequency than all other results---marking it as the solution.
	Since each qubit was chosen to encode one node of the graph, the value of each qubit indicates to which partition a particular node belongs.
	In this case, the solution groups $\{0, 2\}$ and $\{1, 3\}$ as shown in \autoref{fig:wf1_7}, which, indeed, are a solution to the MaxCut problem on this graph.
\end{example}

\section{General Idea}\label{sec:idea}

The established workflow described in \autoref{sec:flow} and illustrated in \autoref{fig:workflow} comes with a major bottleneck:
The crucial and expensive compilation step is conducted \emph{from scratch} during \emph{runtime} for \emph{every} single problem instance---leading to long runtimes for determining a solution for a particular problem instance.
In the following, we propose a \emph{pre-compilation} technique that aims to drastically reduce the time spent on compilation at runtime while maintaining comparable solution quality. This is illustrated in \autoref{fig:workflow2} and described next.

\subsection{Motivation}
For a particular problem class and a selected quantum algorithm, the process of taking a classical problem description and encoding it as a quantum circuit comes with an inherent structure that is independent of the specific instance to be solved.
This can be turned into a \emph{predictive encoding} that captures the structure of the whole problem class at once in the form of a \mbox{general-purpose} quantum circuit.

\begin{example}\label{ex:pred_encoding}
	In case of the MaxCut problem to be solved with QAOA (as considered throughout \autoref{sec:flow}), the problem can be generalized to a problem class by making some kind of assumption on the properties of the graphs that are expected as input.
	The simplest case, essentially assuming that there are no restrictions, is to consider the \emph{complete} graph as shown in \autoref{fig:wf2_1}.
	Any MaxCut problem on four nodes can be derived from this generalized description.
	The predictive encoding is generated by applying the same encoding process to the generalized description as would have been applied to a specific instance.
	More precisely, a qubit is allocated for each of the nodes and a dedicated gate is added to the circuit for each connection between nodes.
	\autoref{fig:wf2_2} figuratively sketches what such a \mbox{all-to-all} circuit might look like.
\end{example}

Since the predictive encoding is, per definition, independent of the particular problem instance, it can be \emph{pre-compiled}, i.e., determined at \emph{compile-time} as opposed to at runtime once the particular instance is known.
This creates a \mbox{general-purpose}, executable circuit for the problem class considered.

\begin{example}
	Consider again the predictive encoding circuit shown in \autoref{fig:wf2_2}.	
	Using \mbox{pre-compilation}, the entire compilation process is conducted and yields a \mbox{pre-compiled} and \mbox{ready-to-execute} circuit at \mbox{compile-time} as sketched in \autoref{fig:wf2_3}.
	\end{example}

At runtime, once the problem instance is known, the previously compiled circuit needs to be \emph{adjusted} to reflect the actual problem instance before being sent to the device provider for execution.
This is significantly cheaper than a full compilation pass as the necessary adjustments frequently boil down to simple gate removals.

\begin{example} \label{ex:circuit_adjustment}
	Assume that, at runtime, the same problem instance previously considered as the starting point in \autoref{fig:wf1_2} should be solved.
	The respective graph (shown again in \autoref{fig:wf2_4}) misses two edges compared to the complete graph that was used for the predictive encoding (shown in \autoref{fig:wf2_1}).
Therefore, the \mbox{pre-compiled} quantum circuit must be traversed and all compiled quantum gates involved in the anticipated but now \mbox{non-present} edges must be removed as shown in \autoref{fig:wf2_5} (insinuated in red).
Afterwards, the altered compiled quantum circuit can be sent for execution and the solution can be decoded as before---leading to the same histogram and solution---again visualized in \autoref{fig:wf2_6} and \autoref{fig:wf2_7}, respectively.
\end{example}

Following this approach, much of the heavy burden of compilation can be shifted from runtime to compile-time.
Furthermore, heavy and compute-intensive optimizations can be applied at compile-time, where it is not so critical to be fast as this has to be done only once.

However, there is no free lunch: The predictive encoding has to subsume a variety of instances which is likely to increase the compilation overhead as compared to compiling a single problem instance.
As demonstrated by experimental evaluations (summarized later in \autoref{sec:eval}), this can be mitigated to some extent by performing lightweight optimizations after the circuit adjustment at runtime.

\subsection{Related Work}\label{sec:rel_work}
So far, \mbox{pre-compilation} has hardly been explored---presumably, because compiling everything at runtime has been \enquote{good enough} for the scale of problems considered today.
Existing works focus on \emph{Variational Quantum Algorithms}~(VQAs) and mostly rely on the concept of lookup tables that map uncompiled (sequences of) gates to their compiled equivalent.

In~\cite{chong2}, already a decade ago, an approach to \mbox{pre-compile} certain rotation angles of \mbox{single-qubit} gates was proposed. 
Whenever a \mbox{single-qubit} with an arbitrary rotation angle occurs, it is compiled by concatenating a sequence of the \mbox{pre-compiled} rotation angles.

After years of no active research in this domain, further works have examined \emph{pre-compilation} on the \emph{pulse} level---describing how each quantum gate is translated into a sequence of \mbox{electro-magnetic} pulses when executed on the actual quantum device.
In~\cite{partialcompilation2019gokhale}, an approach has been proposed targeting VQAs by exploiting their general circuit structure consisting of parameterized and \mbox{non-parameterized} gates.
The authors \mbox{pre-compile} the \mbox{non-parameterized} gates to the pulse level and optimize them extensively. 
At runtime, only the parameterized gates need to be compiled and stitched together with the \mbox{pre-compiled} pulses.

Similarly, the authors of~\cite{cheng2020accqoc} focus on the pulse level and propose a \emph{static/dynamic} hybrid workflow.
This approach is based on storing \mbox{pre-compiled} pulses for certain groups of quantum gates in a database. 
Whenever a new quantum circuit is compiled, it is decomposed into groups and during the static phase, the database of \mbox{pre-compiled} pulses is searched for suitable pulses. 
If there are no \mbox{pre-compiled} pulses for all groups, the respective pulses are generated in the dynamic phase, before they are concatenated with the pulses of the static phase.

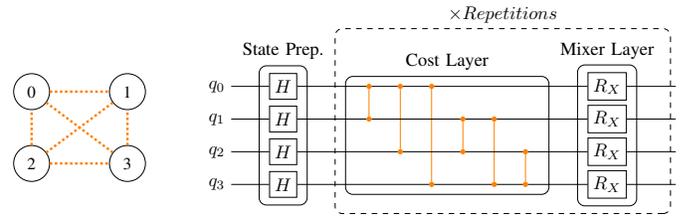
\begin{figure}[t]
				\begin{subfigure}[b]{0.1\textwidth}
				\centering
   \resizebox{0.99\linewidth}{!}{
				\begin{tikzpicture}
    \node[shape=circle,draw=black,minimum size=1.5cm, ] (B) at (0,3) {\huge 0};
    \node[shape=circle,draw=black,minimum size=1.5cm] (D) at (4,3) {\huge 1};
    \node[shape=circle,draw=black, minimum size=1.5cm] (A) at (0,0) {\huge 2};
    \node[shape=circle,draw=black,minimum size=1.5cm] (C) at (4,0) {\huge 3};

    \path [color=orange, dashed, line width =1mm] (A) edge node[] {}(B);
    \path [color=orange, dashed, line width =1mm]  (A) edge node {}(C);
    \path [color=orange, dashed, line width =1mm] (A) edge node[] {}(D);
    \path [color=orange, dashed, line width =1mm] (B) edge node[] {} (C);
    \path [color=orange, dashed, line width =1mm] (B) edge node[] {} (D);
    \path [color=orange, dashed, line width =1mm] (C) edge node[] {} (D); 
				\end{tikzpicture}}
				\vspace{0mm}
				\end{subfigure}
\hfill
\begin{subfigure}[b]{0.35\textwidth}
				\centering
   \resizebox{1.0\linewidth}{!}{
				\begin{tikzpicture}
				  \begin{yquant*}

						qubit {$q_0$} q;
						qubit {$q_1$} q[+1];
						qubit {$q_2$} q[+1];
						qubit {$q_3$} q[+1];

						[this subcircuit box style={rounded corners, inner ysep=4pt,  "State Prep."}]
						subcircuit {
						qubit {} q[4];
				    	box {$H$} q[0-3];
						} (q[0-3]);		
						
						[this subcircuit box style={dashed, rounded corners, inner ysep=4pt,  "$\times Repetitions$"}]
						subcircuit {
						qubit {} q[4];
						
						[this subcircuit box style={rounded corners, inner ysep=4pt,  "Cost Layer"}]
						subcircuit {
						qubit {} q[4];
						[style=orange]
				    	zz (q[0,1]);
						[style=orange]
				    	zz (q[0,2]);
						[style=orange]
				    	zz (q[0,3]);
						[style=orange]
				    	zz (q[1,2]);
						[style=orange]
				    	zz (q[1,3]);
						[style=orange]
				    	zz (q[2,3]);
						} (q[0-3]);			
						
						[this subcircuit box style={rounded corners, inner ysep=4pt,  "Mixer Layer"}]
						subcircuit {
						qubit {} q[4];
						box {$R_X$} q[0-3];
						} (q[0-3]);		
				    	
						} (q[0-3]);

				  \end{yquant*}
				\end{tikzpicture}}
				\end{subfigure}	
				\caption{Predictive Encoding for the MaxCut problem using QAOA assuming that all interactions are present.}
	\label{fig:maxcut}
\end{figure}

\section{Proposed Compilation Scheme}\label{sec:comp_scheme}
In this section, the proposed compilation scheme is explained in more detail---again using MaxCut as a running example.
More specifically, the proposed \mbox{compile-time} and circuit adjustment steps are described.

\subsection{Compile-time Steps}

\begin{figure*}[t]
     \centering
     				\begin{subfigure}[b]{0.10\textwidth}
				\centering
   \resizebox{0.99\linewidth}{!}{
				\begin{tikzpicture}
    \node[shape=circle,draw=black,minimum size=0.5cm, ] (0) at (0,0) {\LARGE 0};
    \node[shape=circle,draw=black,minimum size=0.5cm] (1) at (1,0) {\LARGE 1};
    \node[shape=circle,draw=black, minimum size=0.5cm] (2) at (2,0) {\LARGE 2};
    \node[shape=circle,draw=black,minimum size=0.5cm] (3) at (1,-1) {\LARGE 3};
    \node[shape=circle,draw=black,minimum size=0.5cm] (4) at (1,-2) {\LARGE 4};
    \path [line width =1mm] (1) edge node[] {}(2);   
    \path [line width =1mm] (1) edge node[] {}(3);   
    \path [line width =1mm] (3) edge node[] {}(4);   
    \path [line width =1mm] (0) edge node[] {}(1);    
				\end{tikzpicture}}
		\end{subfigure}
     \hfill
     \begin{subfigure}[b]{0.75\textwidth}
\centering
\resizebox{1.0\linewidth}{!}{
				\begin{tikzpicture}
				  \begin{yquant}
				    	qubit {${q_2}\mapsto Q_0$} q;   
				    	qubit {${q_0}\mapsto Q_1$} q[+1];   
				    	qubit {${q_3}\mapsto Q_2$} q[+1];
				    	qubit {${q_1}\mapsto Q_3$} q[+1]; 	
				    	
				    	box {State \\ Prep.} (-);

						[name=zz1l]
				    	cnot q[3] | q[1];
				    	box {$R_Z$} q[3];
						[name=zz1r]
				    	cnot q[3] | q[1];
				    	
						[name=zz2l]
				    	cnot q[0] | q[1];
				    	box {$R_Z$} q[0];
						[name=zz2r]
				    	cnot q[0] | q[1];
				    	
						[name=out1left]
				    	cnot q[2] | q[1];
				    	box {$R_Z$} q[2];
						[name=out1right]
				    	cnot q[2] | q[1];
				    	
						[name=left]
				    	cnot q[1] | q[3];
				    	cnot q[3] | q[1];
						[name=right]
				    	cnot q[1] | q[3];
				    	
						[name=zz3l]
				    	cnot q[0] | q[1];
				    	box {$R_Z$} q[0];
						[name=zz3r]
				    	cnot q[0] | q[1];
				    	
						[name=out2left]
				    	cnot q[2] | q[1];
				    	box {$R_Z$} q[2];
				    	
						[name=out2right]
				    	cnot q[2] | q[1];
				    	
						[name=left2]
				    	cnot q[1] | q[0];
				    	cnot q[0] | q[1];
						[name=right2]
				    	cnot q[1] | q[0];
				    	
						[name=zz4l]
				    	cnot q[2] | q[1];
				    	box {$R_Z$} q[2];
						[name=zz4r]
				    	cnot q[2] | q[1];

				    	box {Mixer \\ Layer} (-);
				    	
				  \end{yquant}
				  \node[fit=(left)(right), draw, blue, yshift=-0.6cm, inner ysep=20pt, rounded corners] {};
				  \node[fit=(left2)(right2), draw, blue, yshift=0.3cm, inner ysep=12pt, rounded corners] {};
				  \node[fit=(zz1l)(zz1r), draw, black, yshift=0.5cm, inner ysep=20pt, rounded corners] {};
				  \node[fit=(zz2l)(zz2r), draw, black, yshift=-0.2cm, inner ysep=15pt, rounded corners] {};
				  \node[fit=(zz3l)(zz3r), draw, black, yshift=-0.2cm, inner ysep=15pt, rounded corners] {};
				  \node[fit=(zz4l)(zz4r), draw, black, yshift=0.2cm, inner ysep=15pt, rounded corners] {};
				  \node[fit=(out1left)(out1right),  draw, dashed, red, yshift=0.2cm, inner ysep=12pt, rounded corners ] {};
				  \node[fit=(out2left)(out2right), draw, dashed, red, yshift=0.2cm, inner ysep=12pt, rounded corners] {};
				\end{tikzpicture}}
     \end{subfigure}
     \hfill
          \begin{subfigure}[b]{0.1\textwidth}
          
         \centering
   \resizebox{0.99\linewidth}{!}{
				\begin{tikzpicture}
    \node[shape=circle,draw=black,minimum size=1.5cm, ] (0) at (0,3) {\huge 0};
    \node[shape=circle,draw=black,minimum size=1.5cm] (1) at (4,3) {\huge 1};
    \node[shape=circle,draw=black, minimum size=1.5cm] (2) at (0,0) {\huge 2};
    \node[shape=circle,draw=black,minimum size=1.5cm] (3) at (4,0) {\huge 3};

    \path [color=black, draw, line width =1mm] (0) edge node[] {}(1);
    \path [color=black, draw, line width =1mm]  (0) edge node {}(2);
    \path [color=black, draw, line width =1mm] (1) edge node[] {}(2);
    \path [color=black, draw, line width =1mm] (2) edge node[] {} (3);
    \path [color=red, dashed, line width =1mm] (0) edge node[] {} (3);
    \path [color=red, dashed, line width =1mm] (1) edge node[] {} (3); 
				\end{tikzpicture}}          
     \end{subfigure}

	\subfloat[\emph{imbq\_quito}. \label{fig:ibmq_quito}]{\hspace{.10\linewidth}}
	\hspace{2mm}
	\subfloat[Mapped quantum circuit. The state preparation and mixer layer are \mbox{problem-independent} and their compiled gate sequences (as shown in \autoref{ex:nat_gates}) are encapsulated in respective blocks. Each SWAP gate (compiled to three CNOT gates) is denoted in a blue box, each anticipated and also actually present edge is denoted in a black box while the \mbox{to-be-removed} edges are denoted in red boxes. 
	\label{fig:mapping}]{\hspace{.75\linewidth}}
	\hspace{3mm}
	\subfloat[MaxCut problem instance. \label{fig:problem_instance}]{\hspace{.11\linewidth}}
	\caption{Mapping.}
\end{figure*}

The compile-time steps comprise the (predictive) encoding and the (pre-) compilation, which in turn entails native gate-set synthesis and mapping as reviewed in \autoref{sec:flow}.
So far, these steps have mostly been conducted at runtime after the problem instance is known.
In contrast, a different approach has been chosen in the proposed scheme: instead of encoding the actual problem instance, a specific problem instance is anticipated---exploiting the inherent structure of the problem class and the selected quantum algorithm.

\begin{example}\label{ex:enc}
To illustrate the proposed predictive encoding, assume a MaxCut problem with four nodes, and therefore, six possible edges shall be solved using QAOA as already mentioned in \autoref{ex:pred_encoding}.
The general structure of the algorithm (shown in \autoref{fig:maxcut}) consists of a sequence of three different building blocks: the state preparation, the cost layer, and the mixer layer---with the last two blocks being repeated a certain number of times.
To this end, only the cost layer depends on the actual problem instance.
It models the graph by representing each node as a qubit and applying \emph{RZZ} gates between connected nodes, e.g., an edge between nodes $0$ and $1$ translates to a \emph{RZZ} gate between qubits $q_0$ and $q_1$.
By anticipating that all possible edges are present, the cost layer can be fully encoded. 
\end{example}

\vspace{5cm}

After the predictive encoding, all further compilation steps can be conducted in the same fashion as before---just with the difference that the resulting compiled quantum circuit does not represent the actual problem instance (which is not known yet) but an anticipated one.
Using the predictive encoding scheme, \mbox{pre-compilation} is no longer restricted to the \mbox{problem-independent} parts of a general quantum algorithm (as in the existing techniques reviewed in \autoref{sec:rel_work}), but can be applied to the entire quantum circuit.

\begin{example} \label{ex:nat_gates}
Assume that the encoded quantum circuit from \autoref{ex:enc} shall be executed on the five-qubit \emph{ibmq\_quito} device that offers \emph{$R_Z$}, \emph{$\sqrt{X}$}, \emph{X}, and \emph{CX} gates as its native gates and whose layout is shown in \autoref{fig:ibmq_quito}.
Then, each of the QAOA building blocks---state preparation (consisting of \emph{H} gates), cost (\emph{RZZ} gates), and mixer layers (\emph{$R_X$} gates)---is first compiled to native gates using the following circuit identities:
\begin{figure}[h!]
\vspace{-3mm}
\centering
\begin{tikzpicture}
   \begin{yquantgroup}
      \registers{
         qubit {} q[1];
      }
      \circuit{
         h -;
      }
      \equals
      \circuit{
		box {$R_Z(\nicefrac{\pi}{2})$} q[0];
		box {$X$} q[0];
		box {$R_Z(\nicefrac{\pi}{2})$} q[0];
      }
   \end{yquantgroup}
\end{tikzpicture}

\begin{tikzpicture}
   \begin{yquantgroup}
      \registers{
         qubit {} q[2];
      }
      \circuit{
		phase {$\alpha$} q[0] | q[1]; %
        
      }
      \equals
      \circuit{
         cnot q[1] | q[0];
	    box {$R_Z(\alpha)$} q[1];
         cnot q[1] | q[0];
      }
   \end{yquantgroup}
\end{tikzpicture}

\vspace{2mm}

\begin{tikzpicture}
   \begin{yquantgroup}
      \registers{
         qubit {} q[1];
      }
      \circuit{
	    box {$R_X(\alpha)$} q[0];
      }
      \equals
      \circuit{
	    box {$R_Z(\nicefrac{\pi}{2})$} q[0];
	    box {$\sqrt{X}$} q[0];
	    box {$R_Z(\alpha + \pi)$} q[0];
	    box {$\sqrt{X}$} q[0];
      }
   \end{yquantgroup}
\end{tikzpicture}
\vspace{-1mm}
\end{figure}

After that, the circuit's qubits can be assigned to the device qubits as shown on the left-hand side of \autoref{fig:mapping}. 
Consider a single repetition of the cost and mixer layer.
Then, two \emph{SWAP} gates (which are, in turn, decomposed into sequences of three \emph{CX} gates) are necessary to satisfy the connectivity constraints imposed by the architecture throughout the entire circuit.
Eventually, this results in the mapped circuit shown in \autoref{fig:mapping} with the compiled \emph{SWAP} gates denoted in the blue boxes.
\end{example}

After these steps have been executed, a fully executable \mbox{general-purpose} quantum circuit has been created---while the actual problem instance is not known yet.

\subsection{Circuit Adjustment Step}

At runtime, the problem instance is known and the goal is to make sure that the fully encoded, compiled, and mapped circuit is altered such that it actually solves the problem instance.
For that, the difference between the encoded and the actual problem instance must be determined.
Subsequently, the already compiled quantum circuit must be adjusted based on the determined difference.

\begin{example}\label{ex:adjustment}
Consider the same MaxCut instance as in \autoref{ex:circuit_adjustment}.
It comprises four of the six anticipated edges as illustrated in \autoref{fig:problem_instance} (missing edges are denoted as dashed red lines).
Consequently, the \mbox{pre-compiled} quantum circuit (shown in \autoref{fig:mapping}) needs to be adjusted and the (compiled) quantum gates representing the two missing edges must be removed---as highlighted by red boxes each representing one of the to-be-removed edges (while the black boxes represent the actual present edges).
The easiest way to accomplish this---requiring only a linear traversal of the circuit---is by simply setting the angles of all \emph{$R_Z$} gates corresponding to the \mbox{to-be-removed} edges to zero.
This adjustment effectively makes the corresponding \emph{$R_Z$} gates \mbox{no-ops} and hence leads to the applications of the remaining CX gates to cancel each other out.
The resulting circuit is executable and solves the underlying problem---with barely any modifications to the \mbox{pre-compiled} circuit necessary at runtime.
\end{example}

All these benefits do not come for free. 
As stated previously, not only is it important for a circuit to be executable and solve the problem, but the overhead introduced by compilation should be as low as possible.
For most systems, this overhead can be quantified by the number of two-qubit gates in the resulting circuit.
Since predictive encoding anticipates many different problem instances, it is highly likely that its compilation induces a larger overhead compared to compiling a single problem instance.
As such, the proposed \mbox{pre-compilation} scheme offers a trade-off between the time spent during compilation at runtime and the quality of the resulting circuit.
Applying optimizations at runtime allows making this trade-off almost continuous.
The heavier optimizations are applied at runtime, the closer the performance will typically get to the established workflow---at the cost of longer runtimes.

\begin{example}\label{ex:predictive_encoding}
Consider again the circuit from \autoref{ex:adjustment}.
Instead of only setting the angles of the \emph{$R_Z$} gates to zero during circuit adjustment, a lightweight optimization removes these gates entirely from the circuit and tries to cancel subsequent CX gates that act on the same qubits.
Considering \autoref{fig:mapping}, this completely eliminates the gates in the red boxes.
Compared to directly compiling the circuit for the problem instance (as in the established workflow shown in \autoref{fig:workflow}), the proposed scheme results in a circuit containing one more \emph{SWAP} gate.
Note that some parts of the introduced \emph{SWAP} gates can be further canceled with parts of some \emph{RZZ} gates via the lightweight gate optimization---bringing the overhead down to a bare minimum.
\end{example}

As shown by experimental evaluations, which are summarized next, this can reduce the compilation time at runtime by several orders of magnitude while maintaining comparable compiled circuit quality.

\section{Experimental Evaluation}\label{sec:eval}
This section evaluates the proposed compilation scheme for varying problem instances and compares both the compilation time at runtime and the quality of the compiled quantum circuits.
All implementations are available on GitHub (\url{https://github.com/cda-tum/mqt-problemsolver}) as part of the \emph{Munich Quantum Toolkit} (MQT).

\subsection{Setup}
For the evaluation, the MaxCut problem described throughout all examples is implemented based on Qiskit~(v$0.41.1$) in \emph{Python} with various parameters:
\begin{itemize}
\item Number of nodes considered (and therefore qubits): $5$ to $100$ with a step size of $5$.
\item Chosen algorithm: QAOA with $3$ repetitions.
\item Targeted devices: \mbox{ibmq\_quito}, \mbox{ibmq\_montreal}, and \mbox{ibmq\_washington} with $5$, $27$, and $127$ qubits, respectively. Each problem instance is compiled to the smallest but sufficiently large device.
\item Predictive encoding assumption: Edges are possible between all qubit pairs ($e=all$) (as shown in the example in \autoref{ex:predictive_encoding}) or each node can have at most one possible edge to its immediate successor ($e=1$).
\item Problem instance creation: With sample probability \mbox{$p \in \{0.3, 0.7\}$} a possible edge is actually present. The higher $p$, the more anticipated interactions will be present after the problem instance is revealed.
\end{itemize}

All problem instances evaluated are assessed by two criteria: 
\begin{enumerate}
\item Compilation time at runtime and
\item Compiled circuit quality---determined by the number of present \mbox{two-qubit} gates representing the compilation overhead induced by the proposed approach with lower being better.
\end{enumerate}

For comparison, IBM's Qiskit compiler is used with \emph{optimization levels} $O0$ to $O3$ following the quantum workflow (mentioned in \autoref{fig:workflow}) while the \mbox{compile-time} steps of the proposed approach have been conducted with Qiskit's highest optimization level $O3$.
Additionally, it is \mbox{spot-checked} (using \mbox{MQT QCEC}~\cite{pehamEquivalenceCheckingParameterized2023}) that the compiled and adjusted quantum circuits created by the proposed approach are equivalent to the created circuits of the baselines.

\subsection{Results}
\vspace{-1mm}
\newcommand{\picwidth}{0.4}
\begin{figure*}
     \begin{subfigure}[b]{\picwidth\textwidth}
         \centering
         \includegraphics[height=0.75\textwidth]{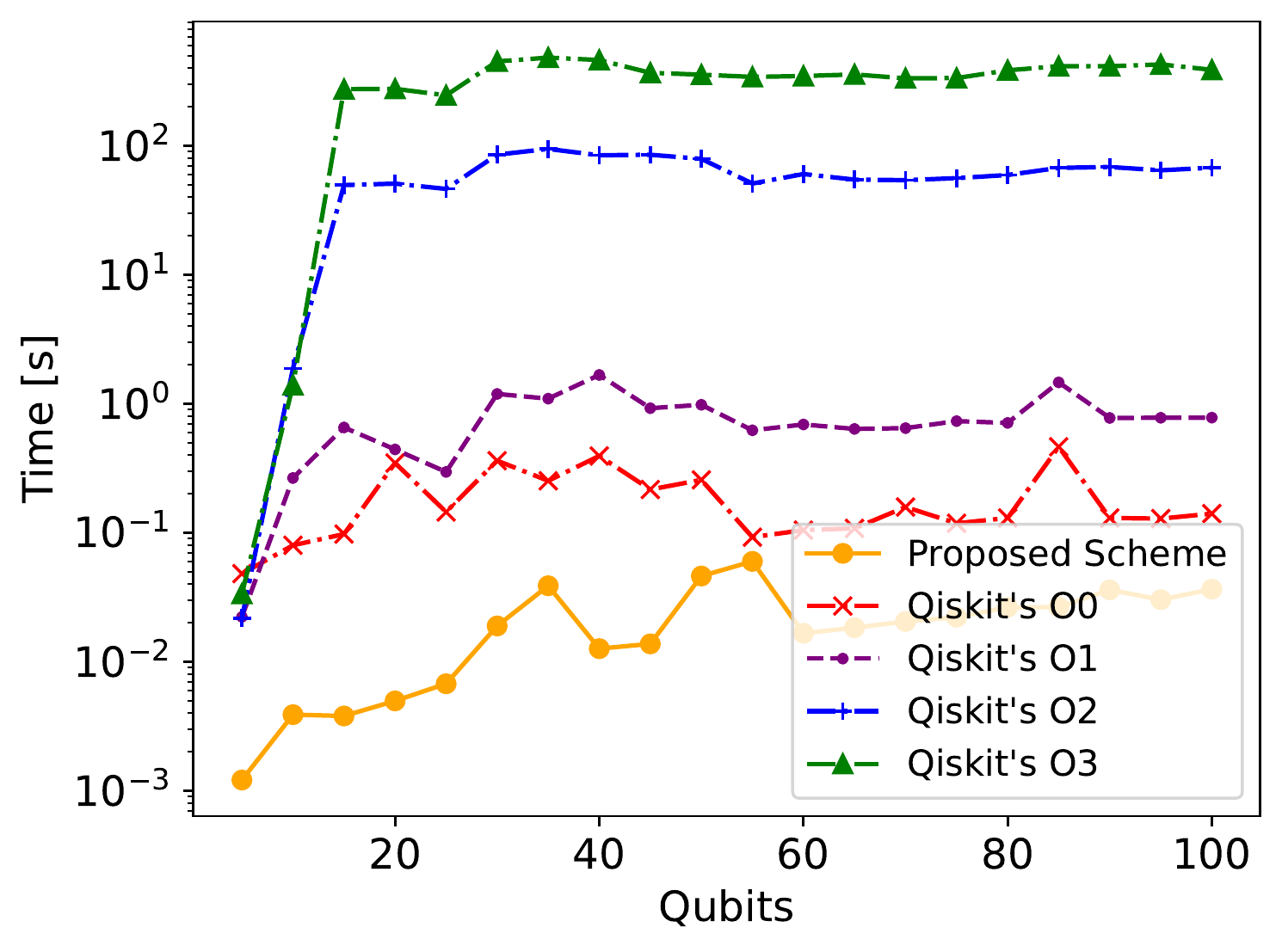}
         \vspace{-7mm}
         \caption{Compilation time at runtime: $e=1$ and $p=0.3$.}
         \label{fig:time_one_03}
         \vspace{3mm}
     \end{subfigure}
     \hfill
     \begin{subfigure}[b]{\picwidth\textwidth}
         \centering
         \includegraphics[height=0.75\textwidth]{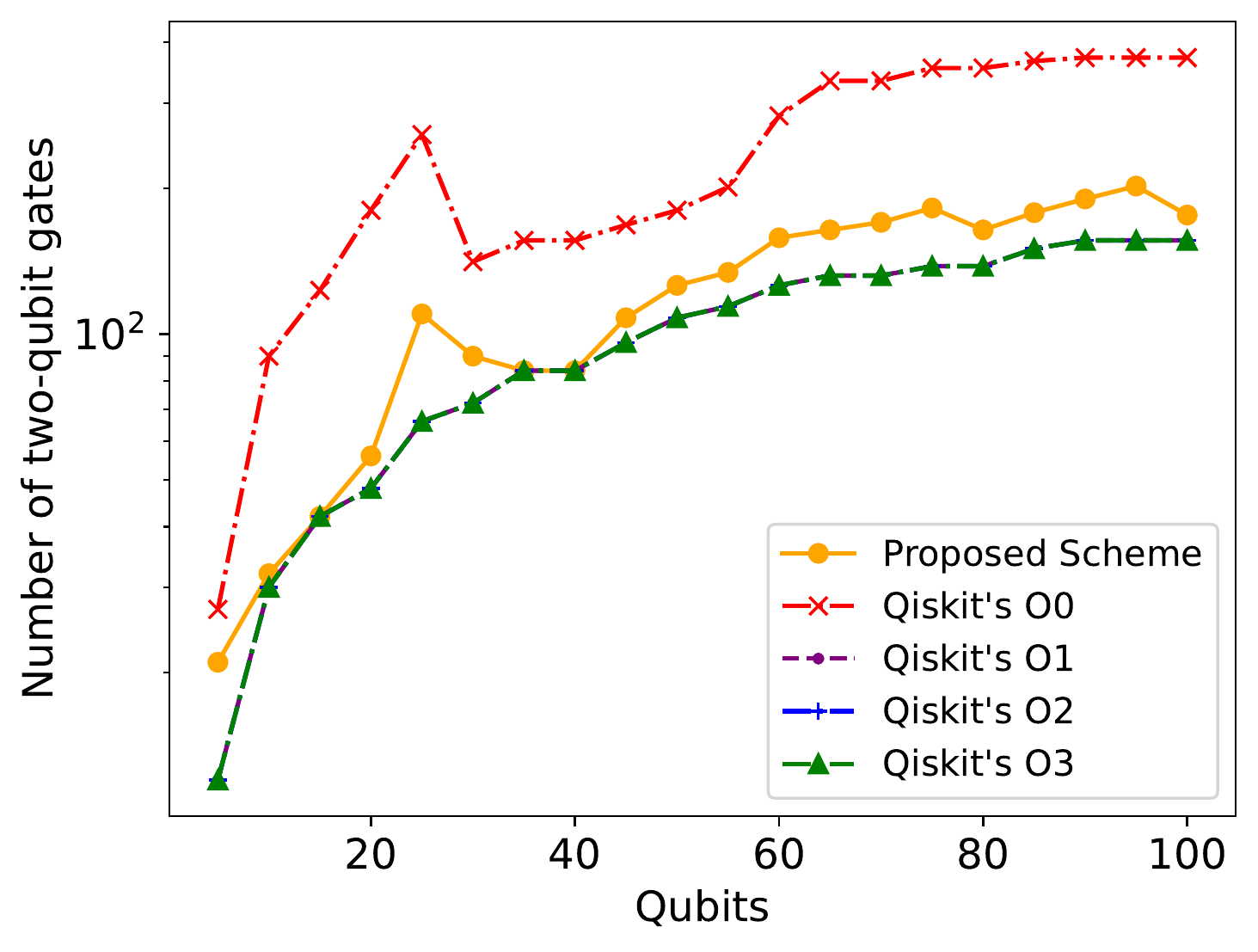}
         \vspace{-7mm}
         \caption{Compiled circuit quality: $e=1$ and $p=0.3$.}
         \label{fig:cx_one_03}
         \vspace{3mm}
     \end{subfigure}
     
                    \begin{subfigure}[b]{\picwidth\textwidth}
         \centering
         \includegraphics[height=0.75\textwidth]{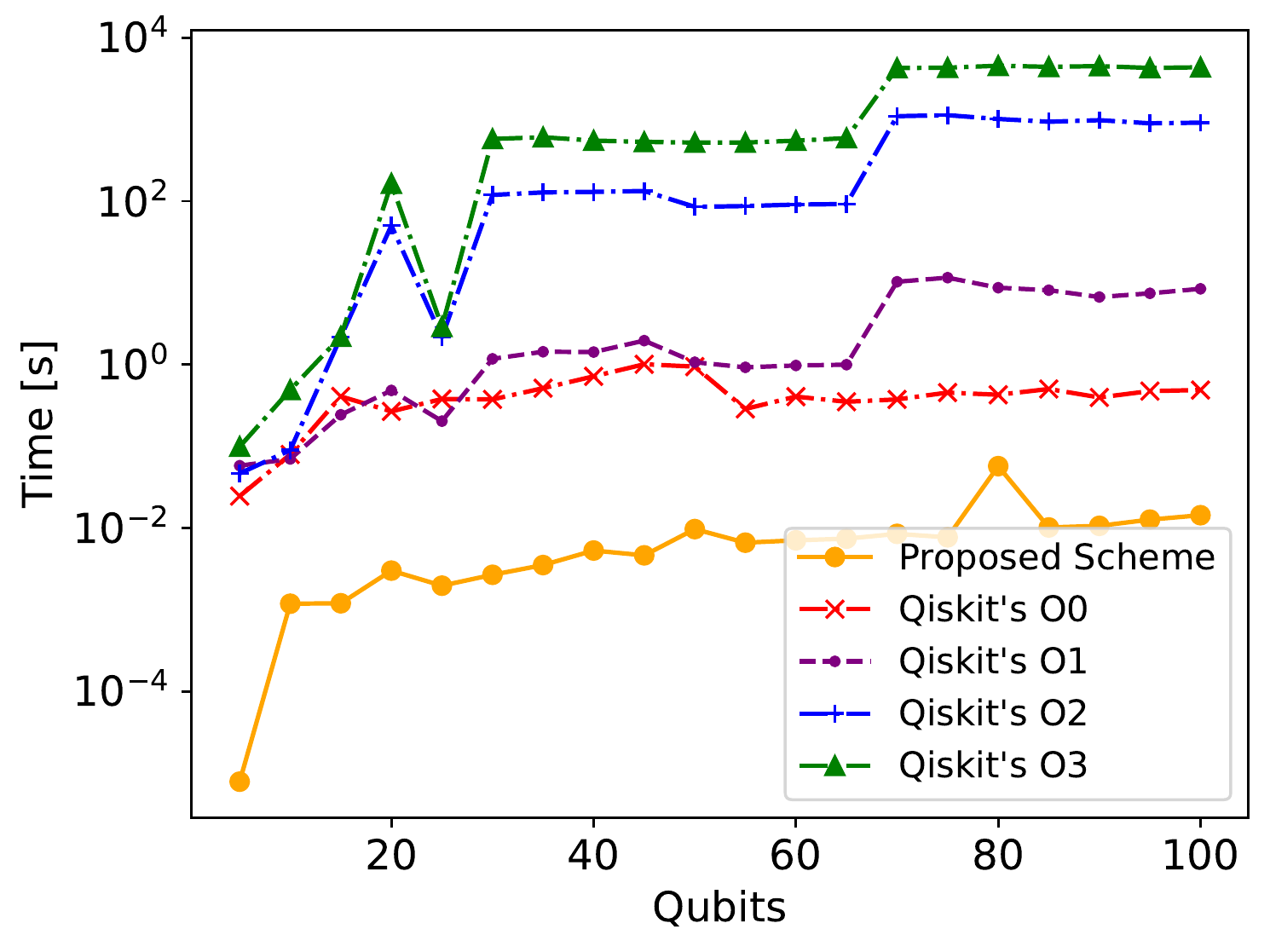}
         \vspace{-7mm}
         \caption{Compilation time at runtime: $e=1$ and $p=0.7$.}
         \label{fig:time_one_07}
         \vspace{3mm}
     \end{subfigure}
     \hfill
          \begin{subfigure}[b]{\picwidth\textwidth}
         \centering
         \includegraphics[height=0.75\textwidth]{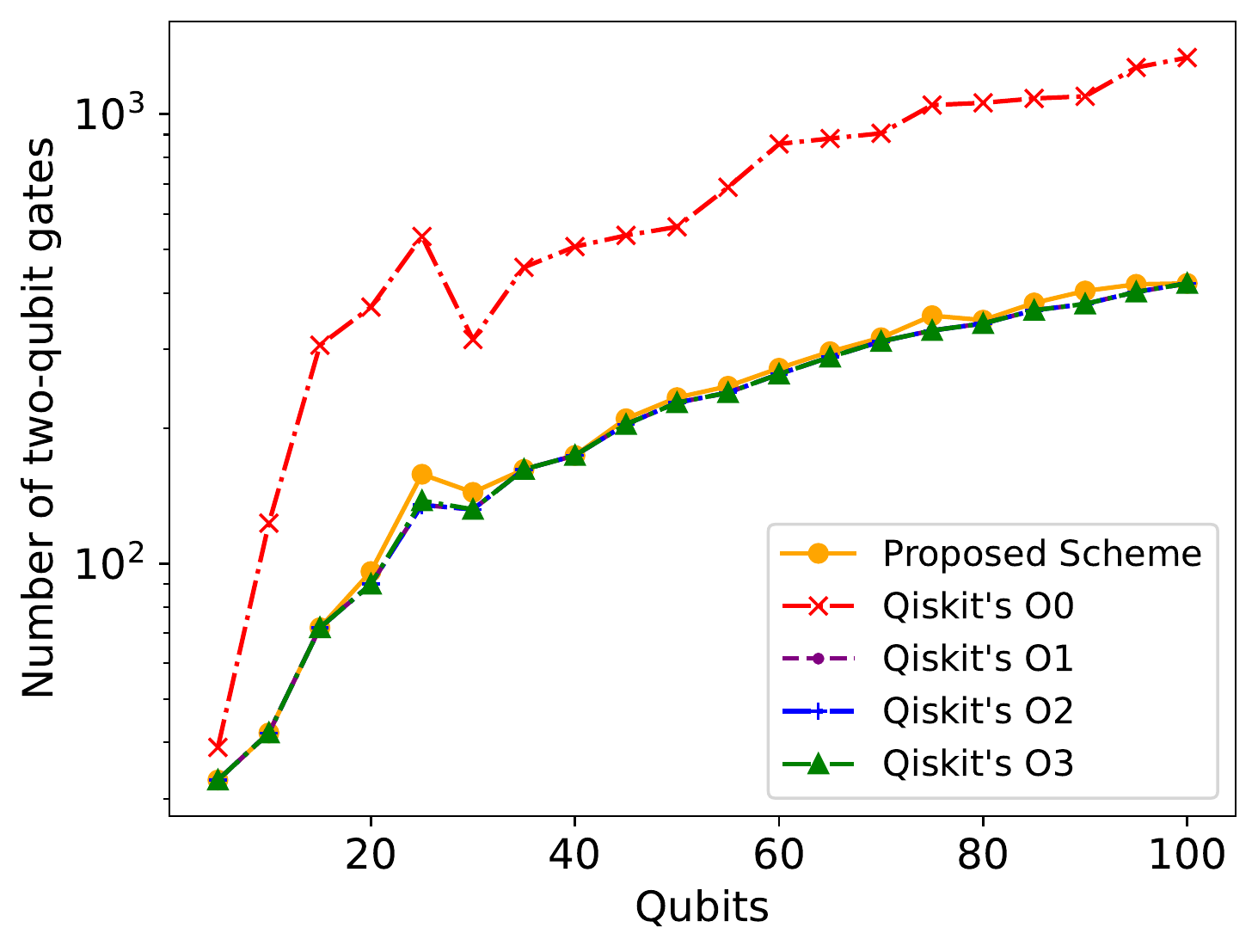}
         \vspace{-7mm}
         \caption{Compiled circuit quality: $e=1$ and $p=0.7$.}
         \label{fig:cx_one_07}
         \vspace{3mm}
     \end{subfigure}
     
     \begin{subfigure}[b]{\picwidth\textwidth}
         \centering
         \includegraphics[height=0.75\textwidth]{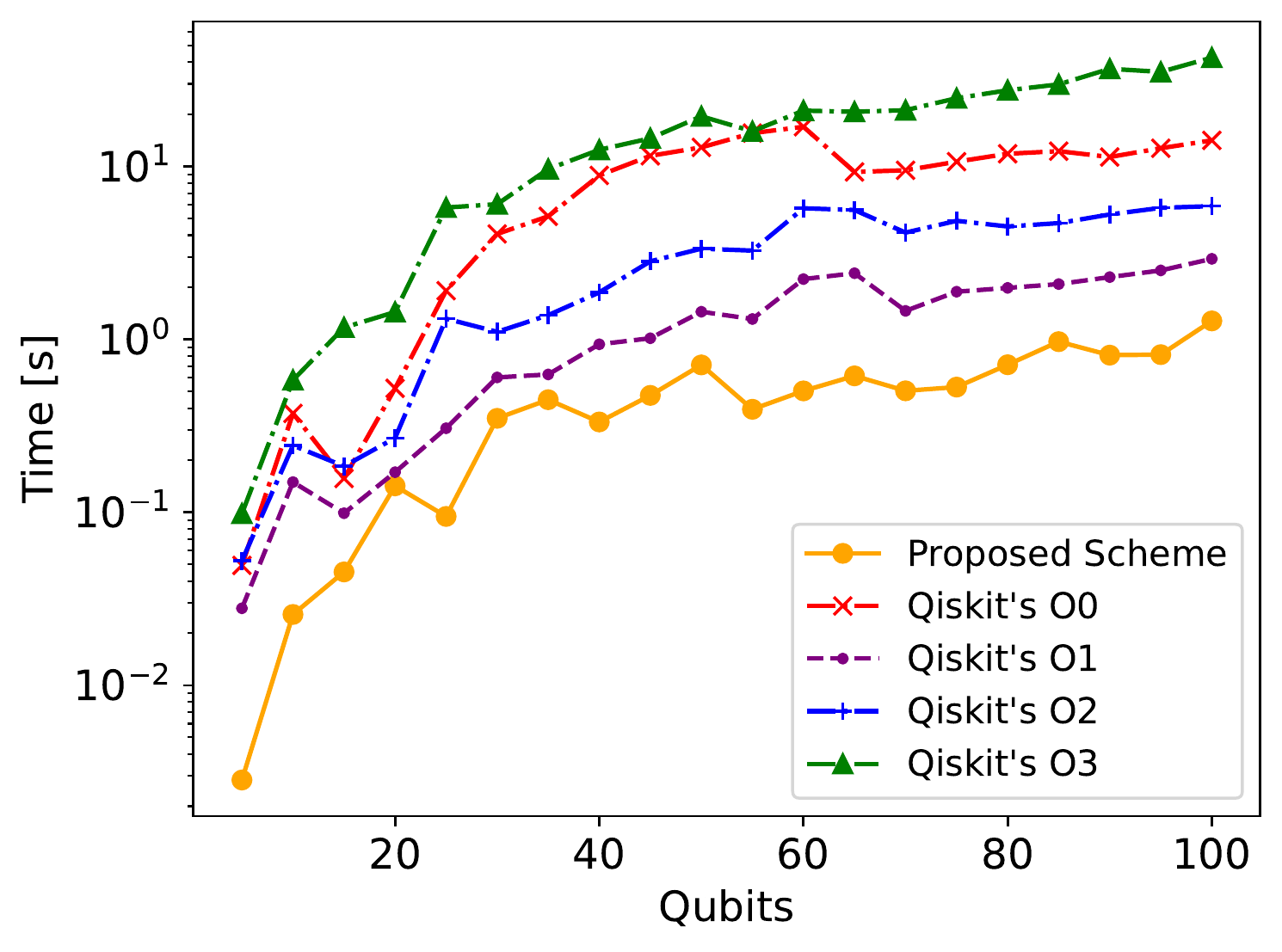}
         \vspace{-7mm}
         \caption{Compilation time at runtime: $e=all$ and $p=0.3$.}
         \label{fig:time_all_03}
         \vspace{3mm}
     \end{subfigure}
     \hfill
               \begin{subfigure}[b]{\picwidth\textwidth}
         \centering
         \includegraphics[height=0.75\textwidth]{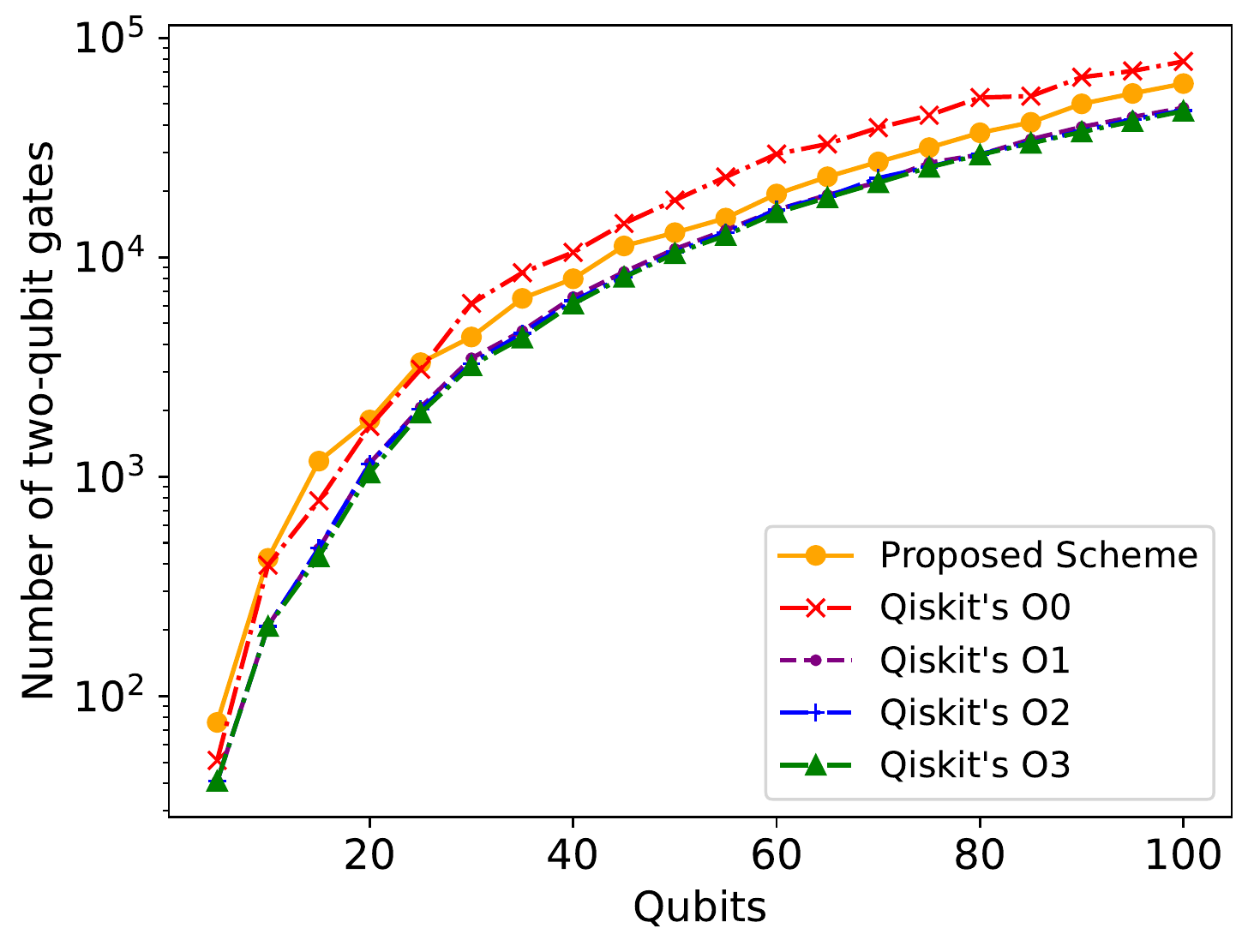}
         \vspace{-7mm}
         \caption{Compiled circuit quality: $e=all$ and $p=0.3$.}
         \label{fig:cx_all_03}
         \vspace{3mm}
     \end{subfigure}
     
          \begin{subfigure}[b]{\picwidth\textwidth}
         \centering
         \includegraphics[height=0.75\textwidth]{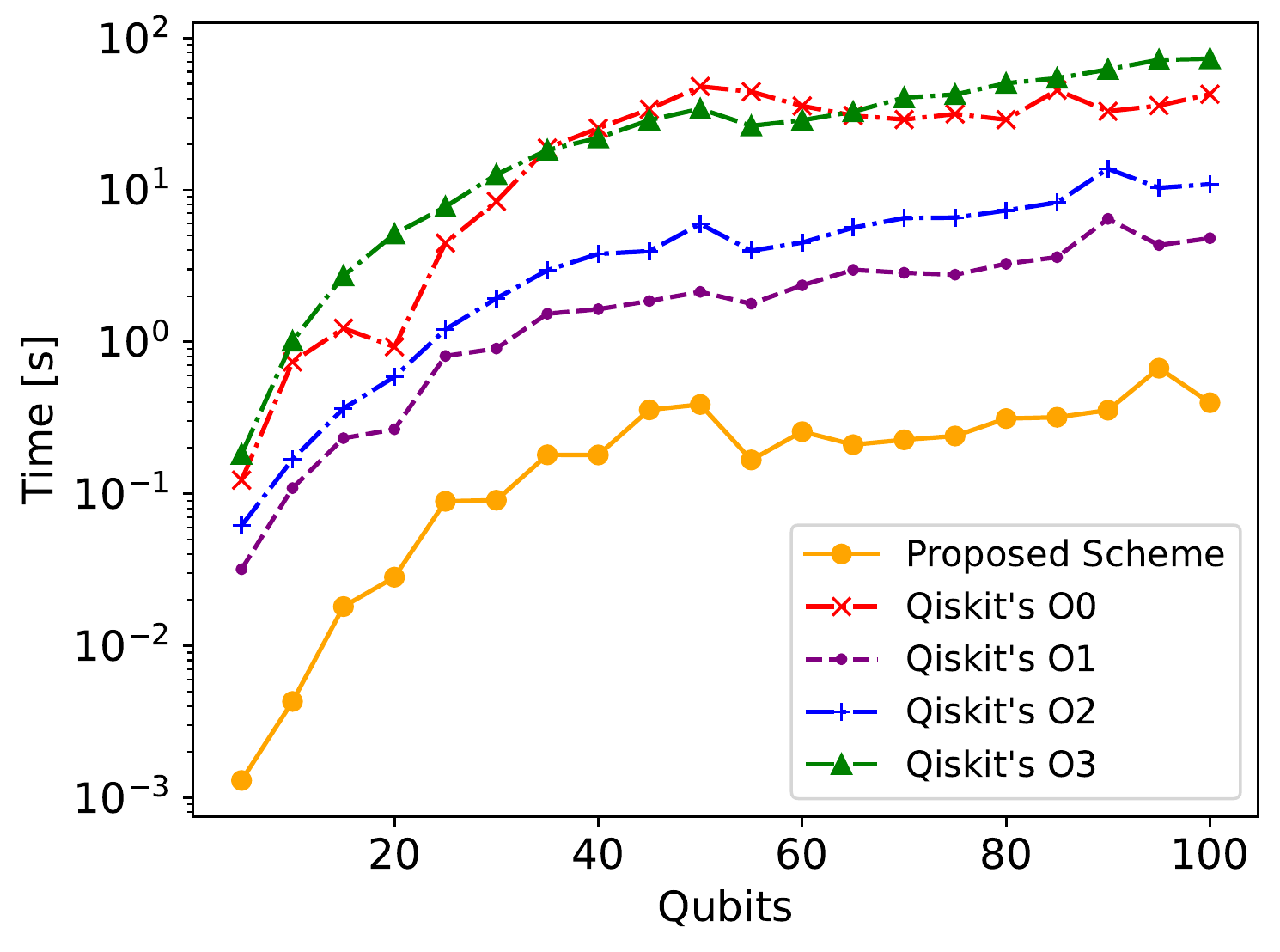}
         \vspace{-7mm}
         \caption{Compilation time at runtime: $e=all$ and $p=0.7$.}
         \label{fig:time_all_07}
         \vspace{3mm}
     \end{subfigure}
     \hfill
        \begin{subfigure}[b]{\picwidth\textwidth}
         \centering
         \includegraphics[height=0.75\textwidth]{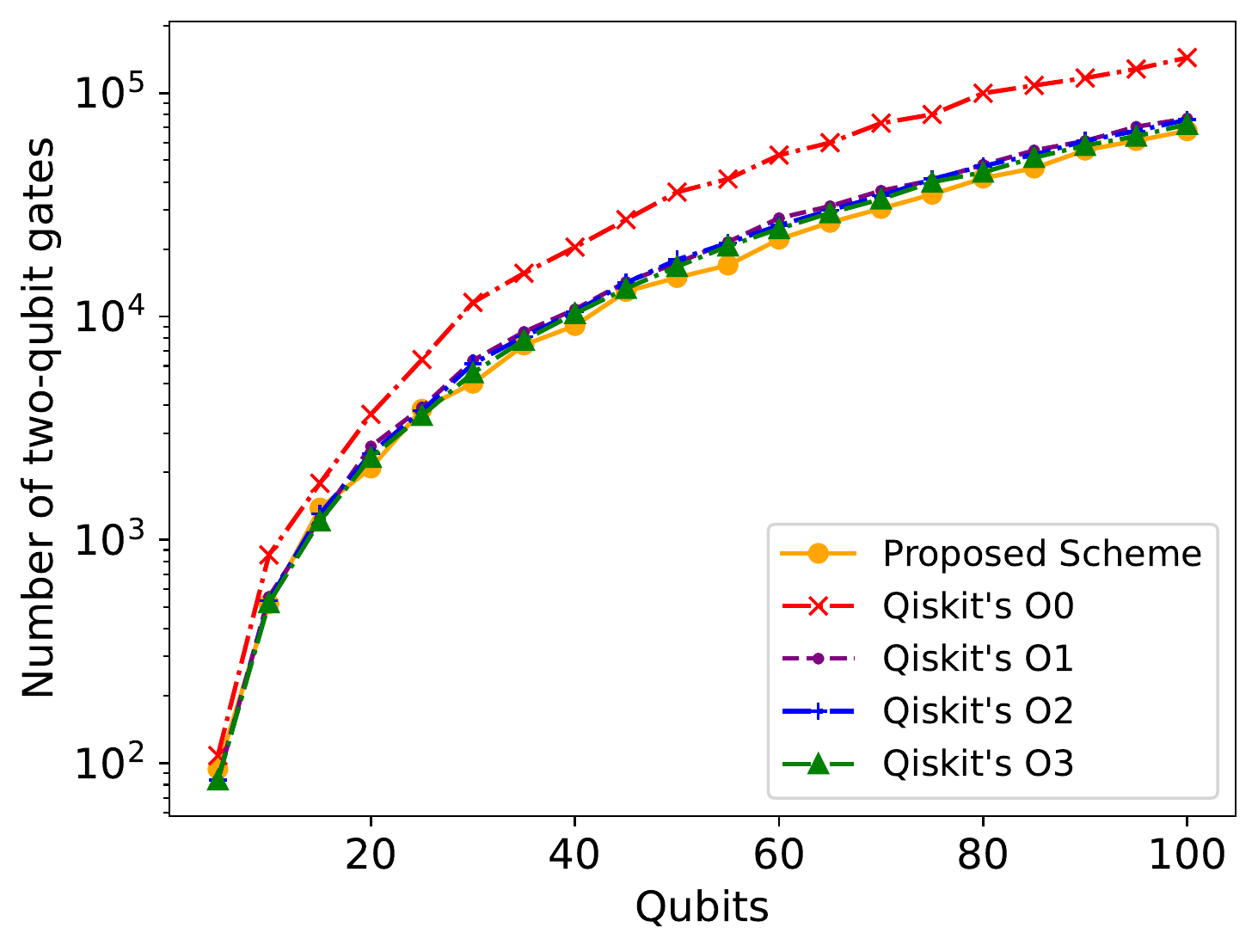}
         \vspace{-7mm}
         \caption{Compiled circuit quality: $e=all$ and $p=0.7$.}
         \label{fig:cx_all_07}
         \vspace{3mm}
     \end{subfigure}
     \caption{Experimental evaluation of the proposed approach for various parameter choices.}
     \label{fig:eval}
\end{figure*}

\begin{figure}[t]
\centering
	\includegraphics[width=0.99\linewidth]{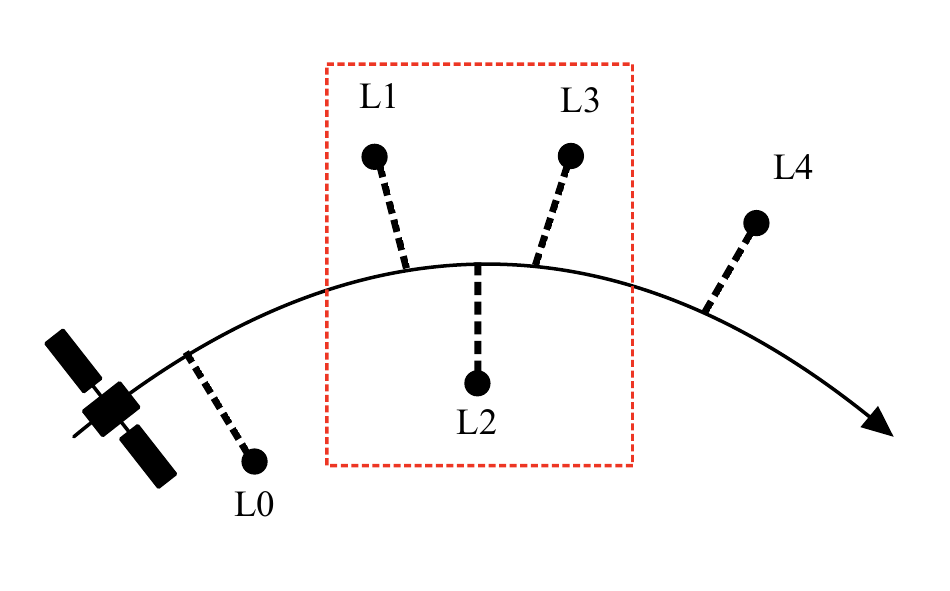}
	\caption{Satellite mission planning problem with five to be imaged locations ($L0$ to $L4$) with two infeasible location pairs: $L1$/$L2$ and $L2$/$L3$.}
	\label{fig:satellite_problem}
\end{figure}

The findings of this study are summarized in \autoref{fig:eval}.
For each combination of $e$ and $p$ values---four combinations in total---the compilation time at runtime and the resulting compiled circuit quality have been evaluated and denoted \mbox{side-by-side}, e.g., in \autoref{fig:time_one_03} and \autoref{fig:cx_one_03} for the combination $e=1$, $p=0.3$.

The results show that the proposed compilation scheme is faster than all Qiskit compilations for all the combinations evaluated of $e$ and $p$ as indicated in the left figures of \autoref{fig:eval}.
When comparing the time reduction depending on the sample probability $p$, it is noteworthy that the improvement is higher for both $p=0.7$ cases compared to $p=0.3$ cases.
The higher $p$ is, the closer the predicted encoding is to the actual problem instances, and consequently, less compilation overhead is induced.

The respective qualities of the compiled circuits are indicated on the right of \autoref{fig:eval}.
The proposed approach, again, outperforms Qiskit's $O0$ and is on par or only slightly worse than the most optimized Qiskit compilation $O3$ in most cases.
When again comparing the $p=0.7$ cases to the $p=0.3$ cases, $p=0.7$ leads to a smaller overhead in terms of the number of \mbox{two-qubit} gates and even outperforms Qiskit's $O3$ for many evaluated problem instances for $e=all$ (as shown in \autoref{fig:cx_all_07}). 

The evaluations demonstrate that the proposed compilation scheme achieves a promising \mbox{trade-off} between compilation time and compiled circuit quality. 
For scenarios where the compilation time at runtime is of prime importance, Qiskit's fastest compilations are the baselines to beat---which the proposed approach is strictly doing for all problem instances.
On top of that, the compiled circuit quality is always better than Qiskit's $O0$ and in the vast majority of cases comparable to Qiskit's most optimized $O3$ compilation---which comes with an inferior compilation time by several orders of magnitude and, by that, is often not a viable option in those scenarios.

\vspace{-1mm}
\section{Application: \\Satellite Mission Planning Problem}\label{sec:satellite}
So far, the benefits of the proposed compilation scheme have been shown and evaluated based on the rather academic MaxCut problem.
In this section, an application from the space industry (taken from~\cite{stollenwerk2020image}) is introduced and the proposed compilation scheme is applied to highlight the benefit it offers for a \mbox{real-world} \mbox{use-case}.

\vspace{-1mm}
\subsection{Motivation}
Currently, there are hundreds of satellites orbiting the Earth with the goal of photographing certain locations.
However, each satellite usually has only a very narrow \mbox{field-of-view} and, therefore, needs to physically rotate its optics between capturing different locations while it is moving on its orbit with constant speed.
As a consequence, it may not be possible to take images of all \mbox{to-be-captured} locations.
Additionally, the time to determine which locations to select is critical and must stay within a fixed time budget---otherwise the result is worthless, since the satellite may already passed the first location.

\begin{example}
Consider a satellite that is tasked to take photos of five locations ($L0$ to $L4$) as shown in \autoref{fig:satellite_problem}. 
While the locations $L0$ and $L1$ are distant and provide enough time for the physical rotation of the optics, there is a selection problem for locations $L1$ to $L3$: If the location $L1$ is selected, it is infeasible to also select the location $L2$. 
Furthermore, if location $L2$ is selected, location $L3$ becomes infeasible.
Therefore, the maximum number of feasible locations is achieved by selecting all except $L2$---resulting in four captured locations.
\end{example}

\subsection{Quantum Computing Approach}
In~\cite{stollenwerk2020image}, this problem has already been approached using a quantum annealing ansatz.
Therefore, a problem formulation as a \emph{Quadratic Unconstrained Binary Optimization} (QUBO) problem was proposed.

In addition to the quantum annealing approach, this problem can also be solved using \mbox{gate-based} quantum computing, e.g., using QAOA.
For that, a simplified problem formulation compared to~\cite{stollenwerk2020image} has been chosen where each location can be captured from one position on the orbit. 
By that, each location is encoded as one qubit representing if that location has been selected for capturing or not---with the goal of capturing as many locations as possible while still being physically feasible.

\begin{figure}[t]
\centering
   \resizebox{1.0\linewidth}{!}{
				\begin{tikzpicture}
				  \begin{yquant*}

						qubit {$q_0$} q;
						qubit {$q_1$} q[+1];
						qubit {$q_2$} q[+1];
						qubit {$q_3$} q[+1];
						qubit {$q_4$} q[+1];

						[this subcircuit box style={rounded corners, inner ysep=4pt,  "State Prep."}]
						subcircuit {
						qubit {} q[5];
				    	box {$H$} q[0-4];
						} (q[0-4]);		
					
					[this subcircuit box style={dashed, rounded corners, inner ysep=4pt,  "$\times Repetitions$"}]
						subcircuit {
						qubit {} q[5];
						
						[this subcircuit box style={rounded corners, inner ysep=4pt,  "Weight Factor"}]
						subcircuit {
						qubit {} q[5];
				    	box {$R_Z(\theta_0\cdot 2\alpha_i$)} q[0];
				    	box {$R_Z(\theta_1\cdot 2\alpha_i$)} q[1];
				    	box {$R_Z(\theta_2\cdot 2\alpha_i$)} q[2];
				    	box {$R_Z(\theta_3\cdot 2\alpha_i$)} q[3];
				    	box {$R_Z(\theta_4\cdot2 \alpha_i$)} q[4];
						} (q[0-4]);

						[this subcircuit box style={rounded corners, inner ysep=12pt,  "Cost Layer"}]
						subcircuit {
						qubit {} q[5];
						[style=black]
						
						phase {$\gamma \cdot 2\alpha_i$} q[1] | q[2];
						[style=black]
						phase {$\gamma \cdot 2\alpha_i$} q[2] | q[3];
						} (q[0-4]);			
						
						[this subcircuit box style={rounded corners, inner ysep=4pt,  "Mixer Layer"}]
						subcircuit {
						qubit {} q[5];
				    	box {$R_X(2 \beta_i)$} q[0-4];
						} (q[0-4]);		
				    	
						} (q[0-4]);

				  \end{yquant*}
				\end{tikzpicture}}
	\caption{Satellite problem encoded using QAOA.}
	\label{fig:satellite_problem_encoding}
\end{figure}
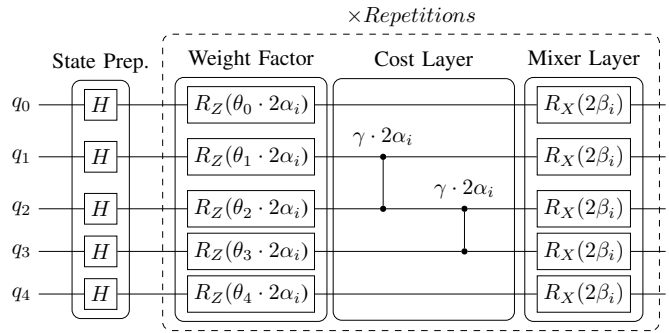

Since QAOA is used---similar as in \autoref{fig:maxcut}---also the circuit structure for the satellite mission planning problem is similar: Whenever two locations cannot be feasibly selected, there is a \emph{RZZ} gate (similar to the edges between nodes in the MaxCut problem). 
The main difference to the MaxCut QAOA example is a weight factor for every qubit (encoded by a \emph{$R_Z$} gate) to 
represent the dependencies between selecting varies locations as shown in \autoref{fig:satellite_problem_encoding} in its uncompiled form for the problem instance depicted in \autoref{fig:satellite_problem}.
Additionally, all \emph{RZZ} gates in the problem layer are multiplied by a factor $\gamma$ determined by the specific problem instance and its translation to an \emph{Ising Hamiltonian}.

\begin{figure*}
     \centering
     \begin{subfigure}[b]{\picwidth\textwidth}
         \centering
         \includegraphics[width=\textwidth]{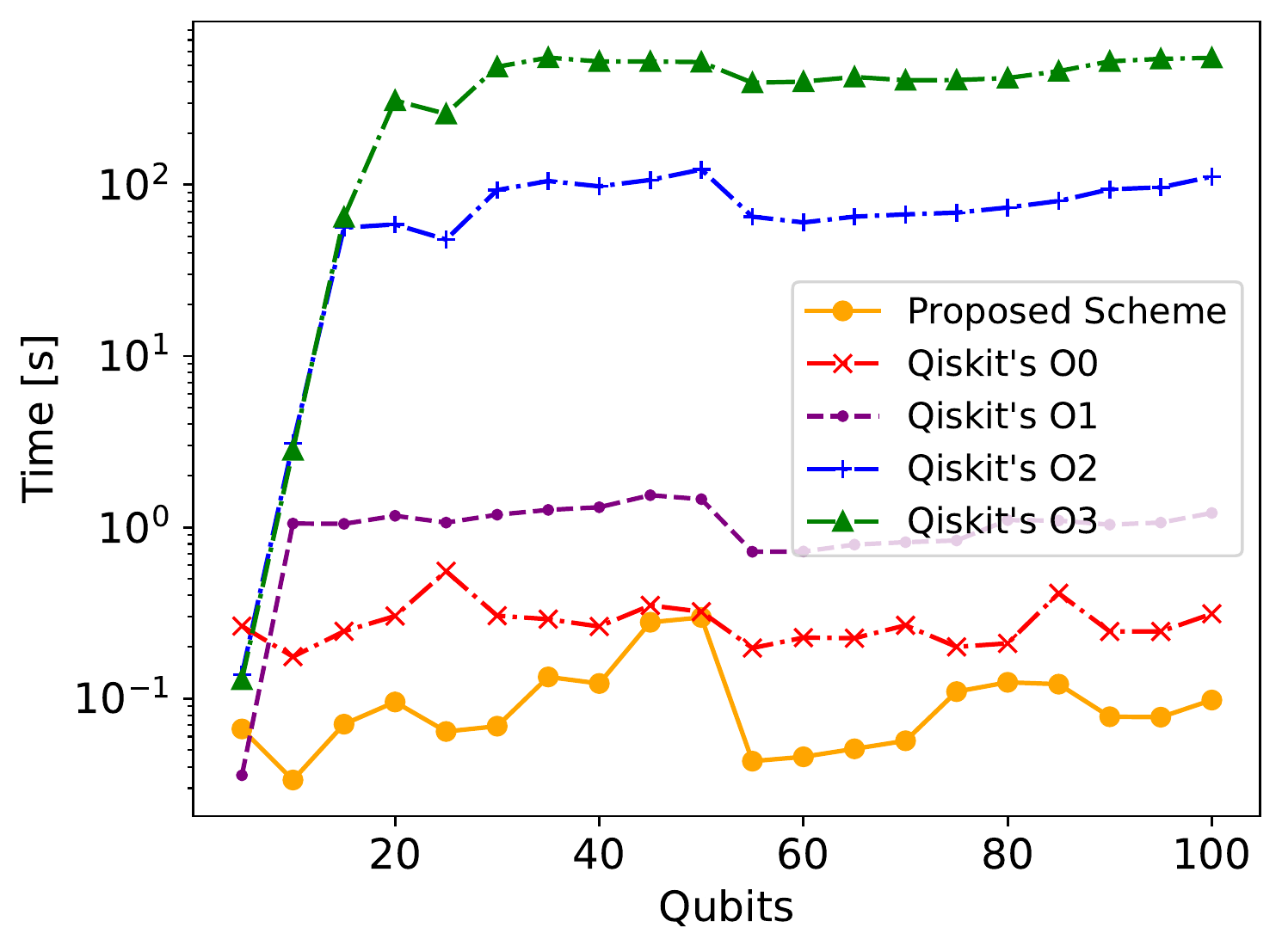}
         \caption{Compilation time at runtime: $p=0.4$.}
         \label{fig:sat_time}
     \end{subfigure}
     \hfill
     \begin{subfigure}[b]{\picwidth\textwidth}
         \centering
         \includegraphics[width=\textwidth]{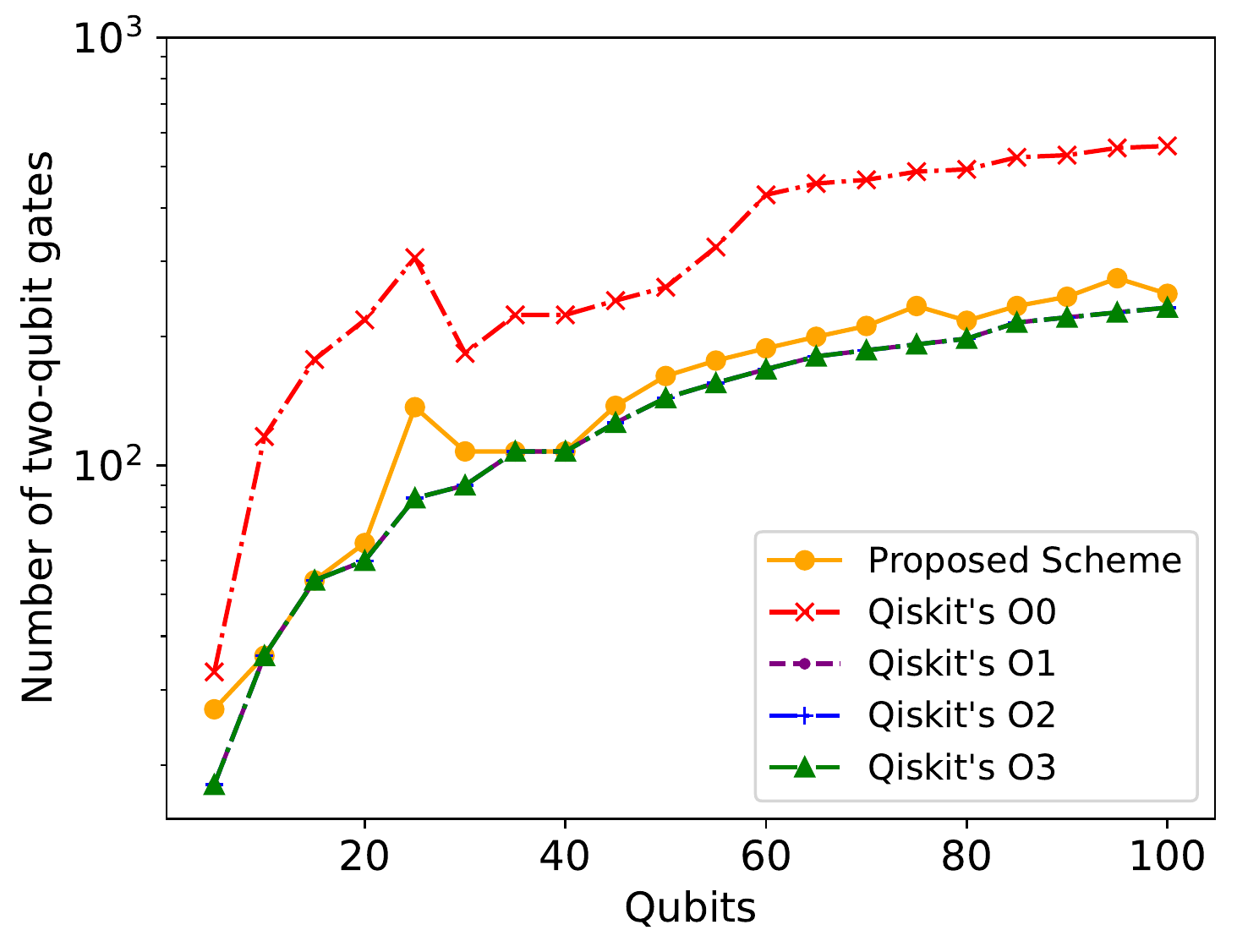}
         \caption{Compiled circuit quality: $p=0.4$.}
         \label{fig:sat_cx}
     \end{subfigure}
\caption{Experimental evaluation for the satellite use-case.}
\label{fig:experiments_sat}
\end{figure*}

\subsection{Experimental Evaluation}
The satellite mission planning problem is evaluated with similar parameters as the MaxCut problem:

\begin{itemize}
\item Number of locations considered (and, therefore, qubits): $5$ to $100$ with a step size of $5$. 
\item Chosen algorithm: QAOA with $3$ repetitions.
\item Targeted devices: \mbox{ibmq\_quito}, \mbox{ibmq\_montreal}, and \mbox{ibmq\_washington} with $5$, $27$, and $127$ qubits, respectively. Each problem instance is compiled to the smallest but sufficiently large device.
\item Predictive encoding assumption: Each pair of subsequent locations is either feasible or infeasible \mbox{($e=1$)}.
\item Problem instance creation: With sample probability \\$p=0.4$ a pair of \mbox{to-be-captured} locations is infeasible.
\end{itemize}

Using the compilation scheme proposed in \autoref{sec:comp_scheme} results in a significantly reduced compilation time at runtime compared to Qiskit's compilation schemes (as depicted in \autoref{fig:sat_time}) while maintaining comparable compiled circuit quality (as depicted in \autoref{fig:sat_cx})---similar to the results of the experimental evaluations described in \autoref{sec:eval}.
The proposed approach is almost always faster than all baselines while the difference becomes even more distinct with an increasing number of qubits.
Regarding the compiled circuit quality, the proposed compilation scheme always outperforms Qiskit's $O0$ and is only slightly worse than Qiskit's most optimized $O3$ compilation---which is inferior by several orders of magnitude in compilation time.

\section{Discussion}\label{sec:discussion}
All those mentioned benefits do not come for free. 
For the predictive encoding, a suitable quantum algorithm and a respective encoding of the problem class must be determined and applied. 
This alone is challenging and requires expert knowledge in both quantum computing and the problem domain---constituting a whole research domain on its own.
The closer the anticipated problem instance comes to the actual problem instance, the better will be the compilation quality.

\vspace{5cm}

Additionally, the encoding scheme must be chosen so that the circuit adjustment step---which adjusts the already fully executable \mbox{pre-compiled} quantum circuit to represent the actual problem instance and not the anticipated one---can be conducted efficiently.
Therefore, it is recommended to only \emph{delete} quantum gates from the \mbox{pre-compiled} circuit.
Inserting new gates could, in the worst case, destroy the mapping so that this compilation step needs to be \mbox{re-done}---impairing the desired compilation time at runtime improvement.
Hence, it heavily influences the \mbox{trade-off} between reduced compilation time and compiled circuit quality---if it is either slow or ineffective, the benefits of the proposed compilation scheme dissolves.

To adapt the proposed scheme for arbitrary problems, some \mbox{one-time} manual effort is required to setup the predictive encoding based on the assumptions made.
The degree of required work needed depends very much on the selected quantum algorithm. 
For some algorithms such as the \emph{Variational Quantum Eigensolver} (VQE,~\cite{peruzzoVariationalEigenvalueSolver2014}) it is simple since the whole quantum circuit is \mbox{problem-independent} and can be \mbox{pre-compiled} already by the approaches mentioned in \autoref{sec:rel_work}.
For other algorithms such as \emph{Grover}~\cite{groverFastQuantumMechanical1996} or \emph{Quantum Phase Estimation} (QPE,~\cite{kitaevQuantumMeasurementsAbelian1995}) it is less \mbox{straight-forward} to find efficient assumptive problem instances, since it is harder to find an overarching anticipated problem instance that can be encoded as a \mbox{general-purpose} circuit where all other possible problem instances can be derived from.

\section{Conclusions}\label{sec:conclusions}
Compilation is an essential---but \mbox{time-consuming}---part of quantum workflows solving problems from any kind of application domain and is currently almost exclusively conducted at runtime.
Additionally, every single problem instance is compiled from scratch, even though they share a similar structure.
This becomes even more severe for frequently recurring problems.

In this paper, we proposed a comprehensive \mbox{pre-compilation} scheme to reduce the compilation time at runtime based on predictive encoding at \mbox{compile-time} and a respective circuit adjustment at runtime.
For that, a \mbox{general-purpose} quantum circuit is created---subsuming a wide variety of problem instances---and \mbox{pre-compiled}.
As soon as the actual problem instance becomes known, the \mbox{pre-compiled} circuit is adjusted to represent the actual problem and not the anticipated one. 

Experimental evaluations on QAOA for the MaxCut problem as well as a case study involving a satellite mission planning problem show that this reduces the compilation time at runtime by several orders of magnitude compared to Qiskit's compilation schemes while maintaining comparable compiled circuit quality.
All implementations are available on GitHub (\url{https://github.com/cda-tum/mqt-problemsolver}) as part of the \emph{Munich Quantum Toolkit} (MQT).

However, implementing this proposed \mbox{pre-compilation} scheme comes with a considerable \mbox{one-time} effort, since a suitable predictive encoding scheme to create the \mbox{general-purpose} circuit is highly dependent on the problem class and the selected quantum algorithm to solve it.
In addition, the encoding scheme must be chosen so that the circuit adjustment step can be conducted efficiently.
Therefore, this work constitutes a first step towards a general comprehensive \mbox{pre-compilation} scheme, but further steps are necessary.

\section*{Acknowledgments}
This work received funding from the European Research Council (ERC) under the European Union’s Horizon 2020 research and innovation program (grant agreement No. 101001318), was part of the Munich Quantum Valley, which is supported by the Bavarian state government with funds from the Hightech Agenda Bayern Plus, and has been supported by the BMWK on the basis of a decision by the German Bundestag through project QuaST, as well as by the BMK, BMDW, and the State of Upper Austria in the frame of the COMET program (managed by the FFG).
\vspace{5cm}

\clearpage
\printbibliography

\end{document}